\documentclass[journal]{IEEEtran}

\usepackage{cite}

\ifCLASSINFOpdf
  \usepackage[pdftex]{graphicx}
\else
  \usepackage[dvips]{graphicx}
\fi

\usepackage{color}
\usepackage{amsmath}

\usepackage[linesnumbered,ruled,commentsnumbered]{algorithm2e}

\ifCLASSOPTIONcompsoc
 \usepackage[caption=false,font=normalsize,labelfont=sf,textfont=sf]{subfig}
\else
 \usepackage[caption=false,font=footnotesize]{subfig}
\fi

\usepackage{booktabs}
\usepackage{multirow}
\usepackage{hyperref}
\usepackage{url}
\usepackage{mathtools}
\usepackage{amssymb}
\usepackage{array,hhline}
\usepackage[table]{xcolor}
\usepackage{cleveref}

\begin{document}

\title{Towards Weakly Supervised Text-to-Audio Grounding}


\author{Xuenan Xu,~\IEEEmembership{Student Member,~IEEE},
Ziyang Ma,~\IEEEmembership{Student Member,~IEEE},
Mengyue Wu,~\IEEEmembership{Member,~IEEE} and
Kai Yu,~\IEEEmembership{Senior Member,~IEEE}
}

\markboth{Journal of \LaTeX\ Class Files,~Vol.~14, No.~8, August~2023}%
{Shell \MakeLowercase{\textit{et al.}}: Bare Demo of IEEEtran.cls for IEEE Transactions on Magnetics Journals}
%

\IEEEtitleabstractindextext{%
\begin{abstract}
Text-to-audio grounding (TAG) task aims to predict the onsets and offsets of sound events described by natural language.
This task can facilitate applications such as multimodal information retrieval.
This paper focuses on weakly-supervised text-to-audio grounding (WSTAG), where frame-level annotations of sound events are unavailable, and only the caption of a whole audio clip can be utilized for training.
WSTAG is superior to strongly-supervised approaches in its scalability to large audio-text datasets.
Two WSTAG frameworks are studied in this paper: sentence-level and phrase-level.
First, we analyze the limitations of mean pooling used in the previous WSTAG approach and investigate the effects of different pooling strategies.
We then propose phrase-level WSTAG to use matching labels between audio clips and phrases for training.
Advanced negative sampling strategies and self-supervision are proposed to enhance the accuracy of the weak labels and provide pseudo strong labels.
Experimental results show that our system significantly outperforms previous WSTAG methods.
Finally, we conduct extensive experiments to analyze the effects of several factors on phrase-level WSTAG.
The code and models are available at \url{https://github.com/wsntxxn/TextToAudioGrounding}.
\end{abstract}

\begin{IEEEkeywords}
text-to-audio grounding, weakly-supervised learning, negative sampling, audio-text representation, clustering
\end{IEEEkeywords}}

\maketitle

\IEEEdisplaynontitleabstractindextext

\IEEEpeerreviewmaketitle

\section{Introduction}

\IEEEPARstart{W}{ith} the development of deep learning and the accessibility of large-scale datasets, audio understanding has achieved remarkable success.
Many works focus on audio understanding tasks such as Acoustic Scene Classification (ASC)~\cite{tan2024acoustic} and Sound Event Detection (SED)~\cite{xu2023semi,gao2024local}, where audio recordings are classified into categories in a closed set, e.g., speech, music.
However, such closed-set classification systems cannot handle complicated requirements like detecting the third beeping sound in an audio recording.
We proposed text-to-audio grounding (TAG)~\cite{xu2021text} to overcome the limitations of SED systems, since TAG aims to detect sound events described by natural language queries.
TAG can be potentially useful in human-machine interaction applications and cross-modal retrieval systems.
It has also contributed to captioning evaluation~\cite{bhosale2023novel}.
In computer vision, visual grounding~\cite{plummer2015flickr30k,gao2017tall} is a fundamental task bridging vision and language.
It is a building block of grounded cross-modal tasks like grounded captioning~\cite{zhou2019grounded} and has been extensively investigated~\cite{zhou2018weakly,wang2021weakly,song2023advancing,huang2023weakly}.
However, as a similar task in the audio domain, TAG has not received as much attention.
In this paper, we address the TAG task to fill in this gap.

\begin{figure}
    \centering
    \includegraphics[width=1.05\linewidth]{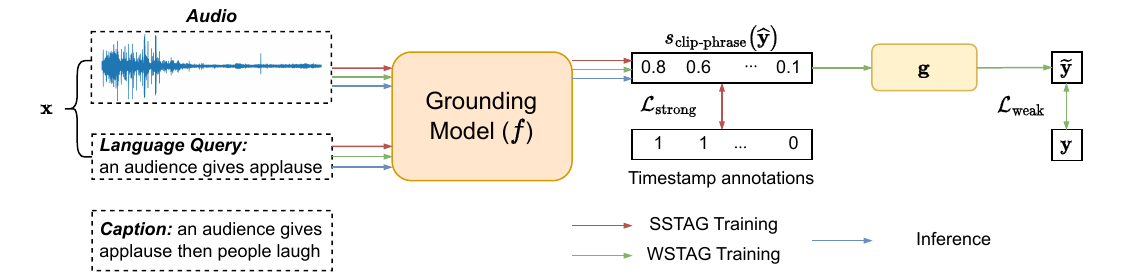}
    \caption{Comparison between SSTAG and WSTAG.}
    \label{fig:strong_weak_compare}
\end{figure}

TAG can be trained in two paradigms: strongly-supervised TAG (SSTAG) and weakly-supervised TAG (WSTAG).
SSTAG uses strong annotations, providing the onsets and offsets of queried events during training.
By contrast, WSTAG only has access to the corresponding caption of each audio while timestamp annotations are unavailable.
\Cref{fig:strong_weak_compare} illustrates the difference between the two paradigms.
Since frame-level supervision signals are provided during SSTAG training, it can achieve better performance than WSTAG.
However, manual strong labeling is expensive and time-consuming, limiting the scalability of SSTAG.
WSTAG can be applied to general audio captioning datasets~\cite{kim2019audiocaps,drossos2020clotho}.
This paper focuses on WSTAG for its scalability.

Here we give a high-level description of weakly-supervised learning to highlight the challenges of WSTAG.
In weakly-supervised learning, the model is trained to predict the desired $\mathbf{\hat{y}}$ from the input $\mathbf{x}$.
However, only pairs of ($\mathbf{x}$, $\mathbf{y}$) are available during training, where $\mathbf{y}$ is the label.
Therefore during training $\mathbf{\hat{y}}$ is fed to an extra module $g$ to obtain $\mathbf{\Tilde{y}}$ so that the model can be trained by minimizing $\mathbf{\Tilde{y}}$ and $\mathbf{y}$.
During inference, the extra module is no longer needed.
The process can be formulated as the following.

\noindent
Training:
\begin{align}
\begin{split}
\mathbf{\hat{y}} = &f(\mathbf{x}), \quad \mathbf{\Tilde{y}} = g(\mathbf{\hat{y}})\\
\ell &= \mathcal{L}(\mathbf{y}, \mathbf{\Tilde{y}})
\label{eq:weakly_supervised_learning}
\end{split}
\end{align}
Inference:
\begin{equation}
\mathbf{\hat{y}} = f(\mathbf{x})    
\end{equation}

Specifically, in WSTAG, $\mathbf{x}$ is (audio, query) while $\mathbf{\hat{y}}$ is the similarity $\mathrm{s}_\mathrm{fp}$ between each audio frame and phrase query.
Previous WSTAG works~\cite{xie2022unsupervised} used the sentence-level audio-caption correspondence as $\mathbf{y}$, which we refer to as \textit{\textbf{sentence-level}} WSTAG.
$g$ includes two steps: pooling along the audio frames to obtain the clip-phrase similarity $\mathrm{s}_\mathrm{cp}$ from $\mathrm{s}_\mathrm{fp}$ and pooling along phrases to obtain the clip-sentence similarity $\mathrm{s}_\mathrm{cs}$.
In previous works~\cite{xie2022unsupervised} mean pooling was used in both steps.
However, this approach violates the multiple instance learning (SMI) assumption of audio description: $\mathrm{s}_\mathrm{cp}$ is high if the phrase occurs in at least one frame, i.e., at least one $\mathrm{s}_\mathrm{fp}$ is high.
According to the definition of weakly-supervised learning, $g$ plays a crucial role since it bridges $\mathbf{\hat{y}}$ and $\mathbf{y}$.
Specifically, in sentence-level WSTAG, the two pooling steps are critical since they bridge the desired fine-grained \textit{frame-phrase} correspondence and the coarse but available \textit{clip-caption} correspondence.
Therefore, in this paper, we explore several pooling strategies and show their influence on sentence-level WSTAG performance.

Furthermore, the training/test mismatch in sentence-level WSTAG hinders performance.
The two pooling steps in $g$ are employed to bridge two natural levels of mismatch: clip/frame in audio and sentence/phrase in text.
In this paper, we attempt to narrow down the mismatch to one level by eliminating the sentence/phrase mismatch.
The phrase-level audio-phrase correspondence is used as $\mathbf{y}$.
$g$ includes only audio pooling.
We refer to this paradigm as \textbf{\textit{phrase-level}} WSTAG.
In phrase-level WSTAG, $\mathbf{y}$ is 1 for positive audio-phrase pairs while 0 for negative ones.
``Positive/negative'' indicates the sound event described by the phrase is present/absent in the audio clip.
To further improve phrase-level WSTAG, we propose two techniques: 
1) advanced negative sampling strategies, including similarity-based and clustering-based, to enhance the accuracy of $\mathbf{y}$;
2) self-supervision, where a pre-trained WSTAG model is utilized to refine $\mathbf{y}$ and provide pseudo labels for $\mathrm{s}_\mathrm{fp}$.
The proposed phrase-level WSTAG, along with the two techniques, brings significant improvement over the sentence-level WSTAG baseline.

In the conference version~\cite{xu2023investigating}, we analyze the influence of several pooling strategies in sentence-level WSTAG where text pooling is used in both training and inference.
This paper uses a simplified sentence-level framework where text pooling is not needed during inference and compares the effects of pooling strategies in the new framework.
Additionally, we extend our work to a new phrase-level WSTAG framework with advanced techniques and analysis.
Our contributions are summarized as follows:
\begin{itemize}
    \item We comprehensively analyze the unique challenge faced in WSTAG: the granularity discrepancy in cross-modal alignment between training and test, and improve WSTAG performance accordingly.
    \item Based on the analysis, we improve sentence-level WSTAG to phrase-level WSTAG to narrow down the training/test discrepancy from the textual modality.
    \item We propose two techniques to improve phrase-level WSTAG, where advanced sampling strategies provide more accurate weak labels and self-supervision further narrows down the discrepancy from the audio modality.
    \item Experiments show our model achieves state-of-the-art (SOTA) WSTAG performance, with comparable performance to SSTAG methods, and generalizes well to SED datasets.
\end{itemize}

\section{Related Work}
\label{sec:related_work}
In this section, we briefly introduce works related to WSTAG from three themes: \textit{weakly-supervised visual grounding}, \textit{weakly-supervised sound event detection}, and \textit{audio-centric text representation learning}.

\subsection{Weakly-supervised Visual Grounding}

Visual grounding is analogous to audio grounding.
It includes two types of tasks: visual object grounding~\cite{plummer2015flickr30k,vasudevan2018object,yang2024exploiting}, and video moment localization~\cite{gao2017tall,hendricks2018localizing,liu2023survey}.
The former requires localizing the objects described by natural language in an image or a video spatially while the latter requires localizing the event described by natural language in a video temporally.
Similar to TAG, high-quality strongly-annotated data with bounding boxes or timestamps is scarce due to expensive labor costs.
Therefore, several works explore weakly-supervised visual grounding approaches~\cite{zhou2018weakly,wang2021weakly,zhang2023cycle} using sentence-level or phrase-level weak labels.
Some works use the sentence-level image/video-sentence correspondence to align the regions in images (or segments in videos) and phrases in sentences by contrastive learning~\cite{datta2019align2ground,da2021asynce,mithun2019weakly}, stimulating a similar setting in WSTAG as described in \Cref{sec:sentence_WSTAG}.
Other works explore using the phrase-level supervision signal for training by reconstructing the phrase queries~\cite{rohrbach2016grounding,liu2019adaptive} or discriminating negative queries from positive ones~\cite{gupta2020contrastive}.
In these works, the phrase-level image/video-phrase correspondence is taken as the training label.
Therefore, phrase-level approaches use more fine-grained visual-text correspondence for training compared with sentence-level ones.
Inspired by these works, we propose phrase-level WSTAG in \Cref{sec:phrase_WSTAG}.

\begin{figure*}[ht]
    \centering    
    \includegraphics[width=0.85\linewidth]{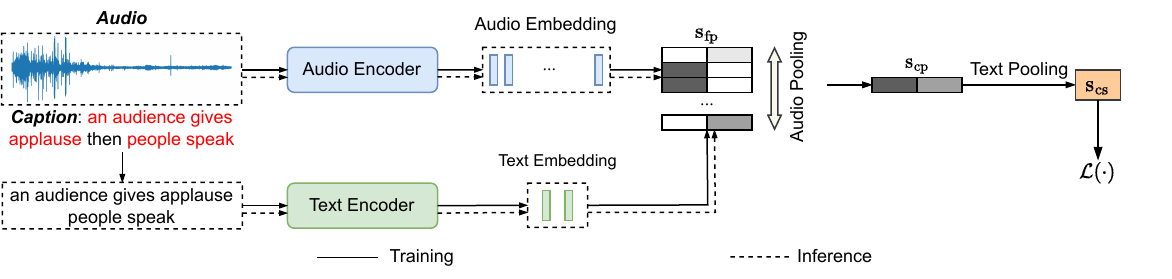}
    \caption{Sentence-level WSTAG. For an audio-caption pair, the frame-phrase similarities $\mathrm{s}_\mathrm{fp}$ are calculated. During training, audio pooling and text pooling transform them into the clip-sentence similarity $\mathrm{s}_\mathrm{cs}$ for loss calculation. During inference, $\mathrm{s}_\mathrm{fp}$ are taken as outputs.}
    \label{fig:sentence_level_WSTAG}
\end{figure*}

\subsection{Weakly-supervised Sound Event Detection}

Weakly-supervised learning has attracted much attention for its potential application on large-scale weakly-annotated data~\cite{zhou2020sal,ren2023weakly,gong2020centroid}.
In the field of SED, weakly-supervised sound event detection (WSSED) has also been explored extensively~\cite{dinkel2021towards,xin2023background,gao2023collecting,li2024weakly}.
It shares a similar goal with WSTAG in detecting onsets and offsets of sound events.
However, in WSSED the sound events are from a pre-defined class set instead of natural language descriptions.
The weakly-annotated data provide only the presence or absence of sound events but the exact timestamps are unavailable. 
According to the weakly-supervised learning paradigm, $\mathbf{\hat{y}}$ and $\mathbf{\Tilde{y}}$ are the estimated frame-level and clip-level event probabilities, respectively.
Current WSSED approaches use neural networks to predict $\mathbf{\hat{y}}$ and then pool it along the temporal axis to obtain $\mathbf{\Tilde{y}}$.
In this work, we adopt the convolutional recurrent neural network (CRNN), which is a popular backbone in WSSED, as the WSTAG audio encoder.
We also investigate the effectiveness of several WSSED temporal pooling strategies in WSTAG.
It should be noted that due to the fixed class set, in WSSED negative (absent) event classes are in fact provided for each data sample. 
In contrast, in WSTAG there is not such an explicit class set so we need to sample negative phrases, i.e., phrases that do not occur in the audio, for each audio sample.

\subsection{Audio-centric Text Representation Learning}
\label{subsec:audio_text_representation}
As a cross-modal task, TAG's performance highly relies on efficient representation learning.
With the development of Transformer-based architectures, large-scale unlabeled data are exploited to extract efficient text ~\cite{devlin2019bert,reimers2019sentence,baevski2022data2vec}, speech~\cite{baevski2019vq, baevski2020wav2vec,hsu2021hubert,ma2023mt4ssl} and visual~\cite{bao2021beit,he2022masked} representations. 
Deep Transformers are trained by masked language modeling (MLM) on the text data, significantly enhancing the performance of downstream language understanding tasks.
Moreover, researchers also investigate learning audio-centric text representation~\cite{vijayakumar2017sound} to facilitate audio-text cross-modal tasks like audio retrieval. 
Recently, contrastive language-audio pre-training (CLAP) has been proposed to extract robust audio and text embeddings from audio-text datasets~\cite{wu2023large,mei2023wavcaps,elizalde2024natural}.
With the large-scale audio-text pre-training, the text encoder is able to extract audio-centric representations from the text.
Phrases describing the same or acoustically similar events are also close in the text embedding space. 
In this work, we use the text representations learned by contrastive pre-training to sample negative phrases in the phrase-level WSTAG.

\section{Sentence-level Weakly-supervised Text-to-audio Grounding}
\label{sec:sentence_WSTAG}

In this section, we first formulate sentence-level WSTAG since it is the basic form of WSTAG and serve as a natural form to investigate the bridge between $\mathbf{\hat{y}}$ and $\mathbf{y}$ via audio and text pooling strategies (see \Cref{eq:weakly_supervised_learning}).
Then we elaborate pooling strategies with regards to their implications in aligning audio and text.

\subsection{Framework}

Sentence-level WSTAG uses sentence-level supervision signals for training.
$\mathbf{\Tilde{y}}$ in \Cref{eq:weakly_supervised_learning} is the clip-sentence similarity $\mathrm{s}_\mathrm{cs}$ and $\mathbf{y}$ is the ground truth clip-sentence correspondence.
As \Cref{fig:sentence_level_WSTAG} shows, for an audio-caption pair $(\mathcal{A}, \mathcal{T})$, two embedding sequences $\{\mathbf{a}_t\}_{t=1}^T$ and $\{\mathbf{t}_n\}_{n=1}^N$ are obtained.
The similarity between the $t$-th audio frame and the $n$-th phrase is calculated as:
\begin{equation}
    \mathrm{s}_\mathrm{fp}(t, n) = \sigma(\mathbf{a}_t \cdot \mathbf{t}_n)
\end{equation},
where $\mathbf{a}_t, \mathbf{t}_n \in \mathbb{R}^e$ and $e$ is the embedding size.
Sigmoid activation restricts $\mathrm{s}_\mathrm{fp}$ to $[0, 1]^{T\times N}$ as it denotes the event probability.
During training, audio pooling and text pooling transform $\mathrm{s}_\mathrm{fp}$ to the clip-sentence similarity $\mathrm{s}_\mathrm{cs}$.
Audio pooling summarizes $\mathrm{s}_\mathrm{fp}$ into the clip-phrase similarity $\mathrm{s}_\mathrm{cp} \in [0, 1]^N$ while text pooling summarizes $\mathrm{s}_\mathrm{cp}$ into $\mathrm{s}_\mathrm{cs}$:
\begin{align}
    \begin{split}
    \small
        \mathrm{s}_\mathrm{cp}(n) &= \mathrm{Pool}_\mathrm{A}\large(\mathrm{s}_\mathrm{fp}(1, n), \mathrm{s}_\mathrm{fp}(2, n), \cdots, \mathrm{s}_\mathrm{fp}(T, n)\large) \\
        \mathrm{s}_\mathrm{cs} &= \mathrm{Pool}_\mathrm{T}\large(\mathrm{s}_\mathrm{cp}(1), \mathrm{s}_\mathrm{cp}(2), \cdots, \mathrm{s}_\mathrm{cp}(N)\large)
    \end{split}
\end{align}
During inference, audio pooling and text pooling are not required.
We use the max margin ranking loss~\cite{karpathy2014deep} for training.
For a minibatch with a batch size $B$, the loss encourages the similarities of positive audio-caption pairs to be higher than any negative pairs by at least the margin $m$:
\begin{align}
    \begin{split}
        &\mathcal{L} = \frac{1}{B} \sum_{i} \sum_{j \neq i}\mathcal{L}_{t}(i, j) + \mathcal{L}_{a}(i, j) \\
        &\mathcal{L}_{t}(i, j) = \max(0, m + \mathrm{s}_{\mathrm{cs}}(i, j) - \mathrm{s}_{\mathrm{cs}}(i, i)) \\
        &\mathcal{L}_{a}(i, j) = \max(0, m + \mathrm{s}_{\mathrm{cs}}(j, i) - \mathrm{s}_{\mathrm{cs}}(i, i))
    \end{split}
\end{align}
where $\mathrm{s}_{\mathrm{cs}}(i, j)$ denotes the similarity between $i$-th audio clip and $j$-th sentence.

In the previous sentence-level WSTAG framework~\cite{xu2023investigating}, the model calculates frame-word similarities.
During inference, the frame-phrase similarities are obtained via another text pooling.
Our new sentence-level WSTAG framework directly calculates frame-phrase similarities so text pooling is no longer needed during inference.
It is more aligned with other methods since all other methods calculate frame-phrase similarities.

\subsection{Pooling Strategies}

\begin{figure*}[ht]
    \centering
    \includegraphics[width=0.85\linewidth]{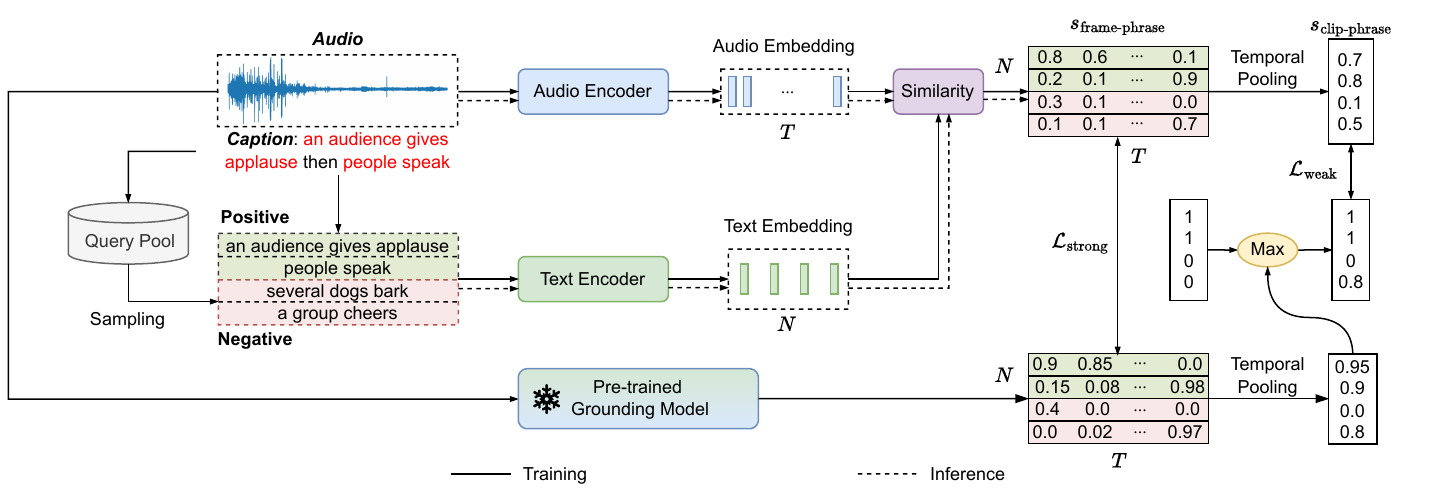}
    \caption{The proposed phrase-level WSTAG approach. For an audio-caption pair, the training data contain both extracted positive phrases and sampled negative ones. A pre-trained WSTAG model is utilized to provide self-supervision. It adopts the same architecture as the WSTAG model to be trained and is trained using only $\mathcal{L}_{\text{weak}}$.}
    \label{fig:sampling_tagging_WSTAG}
\end{figure*}

\paragraph*{\textbf{Audio Pooling}}
\label{subsec:pooling}
Previous work~\cite{xie2022unsupervised} used mean pooling for $\mathrm{Pool}_\mathrm{A}$.
As elaborated in \cite{xu2023investigating}, mean pooling potentially violates the standard SMI assumption of sounds so we investigate temporal pooling strategies utilized in WSSED~\cite{wang2019comparison}, including \textbf{mean pooling}, \textbf{max pooling}, \textbf{linear softmax pooling} and \textbf{exponential softmax pooling}.
\begin{align}
    \footnotesize
    \begin{split}
    \text{Mean:}\quad \mathrm{s}_\mathrm{cp} = \frac{1}{T}\sum_{t}\mathrm{s}_\mathrm{fp}(t)& \quad \text{Max:}\quad \mathrm{s}_\mathrm{cp} = \max\sum_{t}\mathrm{s}_\mathrm{fp}(t)\\
    \text{Linear softmax:}\quad \mathrm{s}_\mathrm{cp} &= \frac{\sum_{t}\mathrm{s}_\mathrm{fp}^2(t)}{\sum_{t}\mathrm{s}_\mathrm{fp}(t)}\\
    \text{Exp. softmax:}\quad \mathrm{s}_\mathrm{cp} &= \frac{\sum_{t}\mathrm{s}_\mathrm{fp}(t)\exp{(\mathrm{s}_\mathrm{fp}(t))}}{\sum_{t}\exp{(\mathrm{s}_\mathrm{fp}(t))}}
    \end{split}
\end{align}
Linear softmax achieves a strong performance in WSSED~\cite{wang2019comparison}.
However, in WSTAG, queries are free text instead of fixed classes and the loss function is also different.
Therefore, linear softmax does not necessarily work well in this scenario.

\paragraph*{\textbf{Text Pooling}}
Text pooling aims to transform $\mathrm{s}_\mathrm{cp}$ into $\mathrm{s}_\mathrm{cs}$.
$\mathrm{s}_\mathrm{cs}$ should be higher if more phrases present high similarity scores with the audio clip so \textbf{sum pooling} seems suitable, which is also used in several weakly-supervised visual grounding works~\cite{karpathy2015deep,zhou2018weakly}.
However, since $\mathrm{s}_\mathrm{cp}$ is bound to be positive, $\mathrm{s}_\mathrm{cs}$ cannot be penalized if there are irrelevant phrases in the caption under sum pooling.
To mitigate this, we also incorporate \textbf{mean pooling} for text pooling.

\section{Phrase-level Weakly-supervised Text-to-audio Grounding}
\label{sec:phrase_WSTAG}


In phrase-level WSTAG, $\mathbf{\hat{y}}$ in \Cref{eq:weakly_supervised_learning} is still the frame-phrase similarity $\mathrm{s}_\mathrm{fp}$ but $\Tilde{\mathbf{y}}$ is the clip-phrase similarity $\mathrm{s}_\mathrm{cp}$.
Phrase-level WSTAG is a combination of SSTAG and WSSED.
SSTAG uses frame-level labels for training.
Phrase-level WSTAG adapts the SSTAG framework to clip-level labels.
As \Cref{fig:sampling_tagging_WSTAG} shows, a training data sample contains an audio clip and $N$ phrases.
Similar to sentence-level WSTAG, $\mathrm{s}_\mathrm{cp}$ is calculated:
\begin{align}
\footnotesize
    \begin{split}
        \mathrm{s}_\mathrm{fp}(t, n) &= \sigma(\mathbf{a}_t \cdot \mathbf{t}_n)\\
        \mathrm{s}_\mathrm{cp}(n) = \mathrm{Pool}\large(\mathrm{s}_\mathrm{fp}(1, n), &\mathrm{s}_\mathrm{fp}(2, n), \cdots, \mathrm{s}_\mathrm{fp}(T, n)\large) 
    \end{split}
\end{align}
Similar to WSSED, the training loss is the clip-level binary cross entropy (BCE) loss between $ \mathrm{s}_\mathrm{cp}$ and the label $y \in \{0, 1\}^N$:
\begin{equation}
\footnotesize
\label{eq:phrase_level_WSTAG_BCE}
    \mathcal{L} = -\frac{1}{N}\sum_{n=1}^Ny(n)\log( \mathrm{s}_\mathrm{cp}(n)) + (1 - y(n))\log(1 -  \mathrm{s}_\mathrm{cp}(n))
\end{equation}

However, phrases from the caption corresponding to the audio (positive phrases) are not enough for training.
Since $y$ for positive phrases is 1, the model will trivially predict probabilities close to 1 if only positive phrases are used for training.
In contrast, in SSTAG there are negative frame-phrase pairs (the phrase is not present in the frame) so strong labels contain 0.
In WSSED, the clip-level labels of an audio clip also contain 0 for absent events.
Therefore, negative phrases, i.e., phrases not present in the audio, are necessary for phrase-level WSTAG.
 
A straightforward approach to negative phrase sampling is to randomly sample phrases from other captions.
However, such phrases are probably not negative for the audio (we call them ``\textbf{false negative phrases}'').
A preliminary analysis shows that over 5\% of all phrases describe the sound of male speech.
If an audio clip contains the sound of a man speaking, phrases describing the same event are likely sampled as negative ones when the number of sampled phrases becomes large.
Therefore, we propose advanced \textbf{sampling strategies} namely similarity- and clustering-based sampling in \Cref{subsec:advanced_sampling}, to ensure that sampled phrases are truly negative.
In addition, to further enhance the label quality and provide frame-level supervision in the weakly-supervised setting, we propose \textbf{self-supervision} where a pre-trained WSTAG model with the same architecture guides training. 

\begin{algorithm}[htpb]
\small
\caption{Similarity-based Negative Sampling.}
\label{alg:similarity_sampling}
\KwIn{Audio captioning data sample $(a, c)\}_{i=1}^N$, query number $n$, qurey pool $Q$,  threshold $\tau$, batch size $b$, similarity function $\mathrm{sim(\cdot)}$.}
\KwOut{WSTAG data sample $(a, q, y)$.}
Extract phrases $\{q_1, q_2, \cdots, q_{n_p}\}$ as $Q_p$ from $c$\;
$Q_n \leftarrow \emptyset$\;
\While{$\lvert Q_n \rvert < n - n_p$}{
Randomly sample $B \subset Q \setminus \{Q_p \cup Q_n\}, \lvert B \rvert = b$\;  
\For{$j=1$ \KwTo $b$}{
$s_j \leftarrow \max_{q}\{\mathrm{sim}(B_j, q)$ for $q$ in $Q_p$\}\;
\If{$s_j < \tau$}{
$Q_n \leftarrow Q_n \cup \{q_j\}$\;
}
\If{$\lvert Q_n \rvert = n - n_p$}{\textbf{break}}
}
}
$q \leftarrow Q_p \cup Q_n$\;
$y \leftarrow [0] * n$\;
$y[:n_p] = 1$\;
\textbf{return} $(a, q, y)$\;
\end{algorithm}

\subsection{Advanced Sampling Strategy}
\label{subsec:advanced_sampling}
\paragraph{Similarity-based sampling}
Similarity-based sampling is illustrated in \Cref{alg:similarity_sampling}.
It calculates the similarities between sampled phrases and positive phrases.
Then phrases that are not negative enough are filtered out by a threshold.
Since the size of the query pool can be very large, we adopt an efficient batch subset selection algorithm.
It keeps sampling negative queries in batches until $n$ queries are sampled.  
We calculate similarities between phrases by transforming phrases into audio-centric text representations introduced in \Cref{subsec:audio_text_representation}.
We first train a bi-encoder using \textbf{C}ontrastive \textbf{L}anguage-\textbf{A}udio \textbf{P}re-training (CLAP) on the WSTAG dataset.
Then the text encoder of CLAP is used to extract audio-centric embeddings.
It should be noted that although we use ``CLAP'', it is \textbf{trained solely on the WSTAG dataset without using extra data}.
Acoustically similar phrase pairs are also close to each other in this text embedding space.
Such a similarity-based sampling prevents phrases describing events in the audio clip from being sampled as negative phrases.

\begin{figure}[ht]
    \centering
    \includegraphics[width=0.95\linewidth]{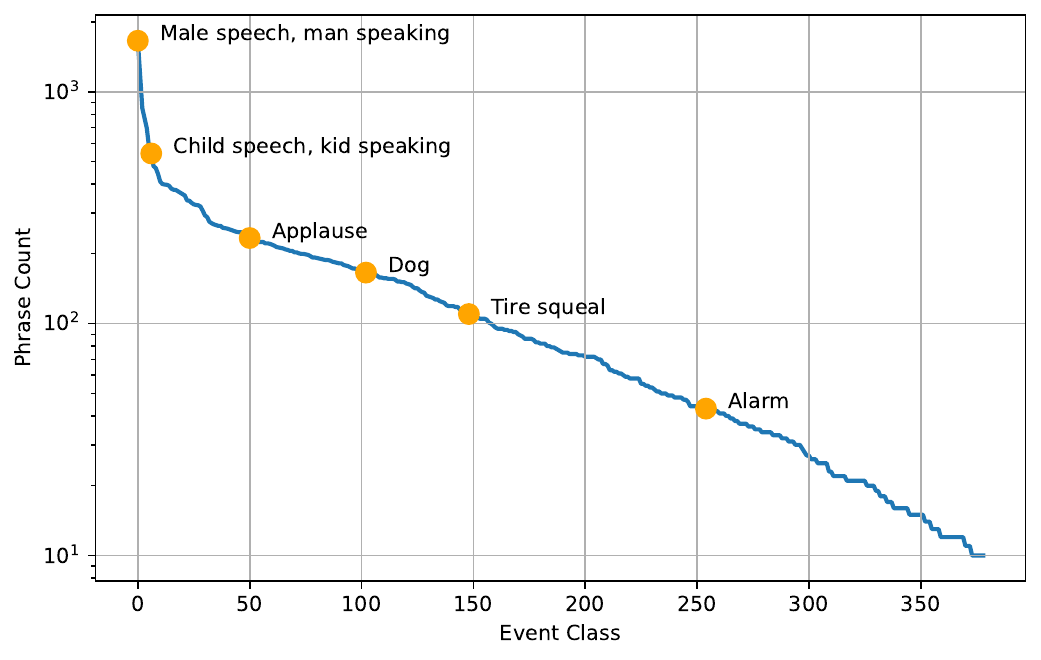}
    \caption{The event distribution of phrases. Each phrase is mapped to its most acoustically similar AudioSet event. The phrase count of each class is plotted.}
    \label{fig:phrase_event_distribution}
\end{figure}

\begin{algorithm}[!ht]
\caption{Clustering-based Negative Sampling.}
\label{alg:clustering_sampling}
\small
\KwIn{Audio captioning data sample $(a, c)$, query number $n$, clusters $C = \{s_i\}_{i=1}^{n_c}$, mapping function $\mathrm{M}(\cdot)$ from phrases to clusters.}
\KwOut{WSTAG data sample $(a, q, y)$.}
Extract phrases $\{q_1, q_2, \cdots, q_{n_p}\}$ from $c$\;
$C_p \leftarrow \cup\{\mathrm{M}(q_1), \mathrm{M}(q_2), \cdots, \mathrm{M}(q_{n_p})\}$\;
$C_n \leftarrow C \setminus C_p$\;
$Q_n \leftarrow \emptyset$\;
$n_{sample} \leftarrow \min(\lvert C_n \rvert, n - n_p)$\;
\For{$j=1$ \KwTo $n_{sample}$}{
  Randomly sample $q_j$ from $C_n[j]$\;
  $Q_n \leftarrow Q_n \cup \{q_j\}$\;
}
\If{$n - n_p > \lvert C_n \rvert$}{
  \For{$j=1$ \KwTo $n - n_p - \lvert C_n \rvert$}{
    Randomly sample $q_j$ from $C_n[j \bmod \lvert C_n \rvert]$\;
    $Q_n \leftarrow Q_n \cup \{q_j\}$\;
  }
}
$q \leftarrow Q_p \cup Q_n$\;
$y \leftarrow [0] * n$\;
$y[:n_p] = 1$\;
\Return $(a, q, y)$;
\end{algorithm}


\paragraph{Clustering-based sampling}
Although similarity-based sampling filters out false negative phrases, the event distribution of sampled phrases is unbalanced.
We map all phrases to AudioSet~\cite{gemmeke2017audio} classes based on the text similarity and plot the event distribution in \Cref{fig:phrase_event_distribution}.
There are over 1,000 phrases mapped to the sound of male speaking while the most infrequent class matches only 10 phrases.
Therefore, it is highly probable that phrases describing infrequent events are seldom sampled by similarity-based sampling.
In contrast, in WSSED, labels of all events are always available.
To make the event distribution more balanced, we propose clustering-based sampling, as \Cref{alg:clustering_sampling} shows.
We first do clustering on all phrases using their CLAP embeddings. 
In this way, phrases belonging to the same sound event or sharing the same acoustic characteristics are grouped into the same cluster.
Negative phrases are sampled from each cluster iteratively to make the event distribution as even as possible.

\subsection{Self-Supervision}
The proposed sampling strategies are effective in achieving high-quality and balanced negative sampling.
However, due to its weakly-supervised nature, noise still exists in clip-phrase matching labels, mostly derived from the following two aspects: 
\begin{itemize}
    \item The text matching or clustering based on CLAP embeddings is not perfect, so false negative phrases are not filtered out or assigned to the wrong groups.
    \item Some sound events are not included in the corresponding caption. Annotators are likely to focus only on prominent events while neglecting background sounds like wind blowing. As a result, phrases describing these events are mistakenly sampled.
\end{itemize}

Self-training and distillation has been found useful in approaches like noisy student training~\cite{xie2020self,park2020improved}.
We adopt a similar approach by taking soft labels estimated by a WSTAG teacher model for supervision and define it as WSTAG self-supervision.
Although noise is inevitable in the training data, the teacher model is able to reasonably estimate clip-phrase similarities, especially for sampled negative phrases.
The sufficient supervision of positive phrases, which are almost accurate, enables the model to recognize sound events when they are present.
Therefore, a pre-trained model G is utilized to refine $y$ in \Cref{eq:phrase_level_WSTAG_BCE}.
Furthermore, in addition to clip-level similarities, G also predicts frame-level ones.
We also employ these frame-level similarities as training labels, thereby mitigating the drawback of WSTAG that frame-level supervision is lacking.
In this way, the weakly-annotated data is leveraged so that G provides self-supervision.

Formally, G estimates the probabilities $y_\text{self}$ of phrases $\mathcal{P}$ given the audio $\mathcal{A}$.
Then the maximum value of $y$ and aggregated clip-level predictions $\hat{y}_\text{self} \in [0, 1]^{N}$ is used as the refined label $y_\text{refined} \in [0, 1]^N$:
\begin{align}
    \small
    \label{eq:self_supervision}
    \begin{split}
        y_\text{self}(t, n) &= \text{G}(\mathcal{A}, \mathcal{P})\\
        \hat{y}_\text{self}(n) &= \text{Pool}_T(y_\text{self}(1, n), y_\text{self}(2, n), \cdots, y_\text{self}(T, n))\\
        y_\text{refined}(n) &= \max(y(n), \hat{y}_\text{self}(n))
    \end{split}
\end{align}
$y_\text{refined}$ replaces $y$ in \Cref{eq:phrase_level_WSTAG_BCE} to calculate the weak loss $\mathcal{L}_\text{weak}$.
The strong loss $\mathcal{L}_\text{strong}$ is calculated on the frame level and the training loss is the combination of these two:
\begin{align}
    \footnotesize
    \begin{split}
    \mathcal{L}_\text{strong}(t, n) &= -y_\text{self}(t, n)\log( \mathrm{s}_\mathrm{fp}(t, n)) - (1 - y_\text{self}(t, n))\log(1 - \mathrm{s}_\mathrm{fp}(t, n))\\
    \mathcal{L} &= \frac{1}{N\cdot T}\sum_{n=1}^N\sum_{t=1}^T\mathcal{L}_\text{strong}(t, n) + \frac{1}N\sum_{n=1}^N\mathcal{L}_\text{weak}(n)
    \end{split}
\end{align}
Here G adopts exactly the same architecture as the target model.
Therefore, the training contains two stages: 1) train a WSTAG model G using \Cref{eq:phrase_level_WSTAG_BCE}; 2) use G as a teacher to train the student model using \Cref{eq:self_supervision}.

\section{Experimental Setup}
\label{sec:exp_setup}

\subsection{Datasets}
In this work, all models are trained on AudioCaps~\cite{kim2019audiocaps}, an audio captioning dataset without frame-level labels.
There are 49501, 2475, and 4820 audio-caption pairs in the training, validation, and test set of our downloaded version, respectively. 
The three sets are merged and re-split into a training set and a validation set with 1,000 audio-caption pairs. 
The audio-centric text encoder used in phrase-level WSTAG is also trained on AudioCaps.
We use AudioGrounding~\cite{xu2021text} for evaluation.
AudioGrounding is a subset of AudioCaps augmented with human-annotated onsets and offsets of phrases.
When training models on AudioCaps, audio files in AudioGrounding test set are eliminated.
We only use the test set for evaluation.
Although all audio-text datasets can be utilized for training, we only use AudioCaps in this work since AudioGrounding is its subset.
Text descriptions from other datasets are found to have a different style from AudioCaps so incorporating these datasets may not be helpful.

\subsection{Model Architectures}
\label{subsec:model_arch}
In this work, we use the same architectures for all approaches.
The audio encoder is a CRNN with 8 CNN layers and a bidirectional GRU (BiGRU).
The CNN structure is similar to CNN10 in PANNs~\cite{kong2020panns} with the modification that the temporal down-sampling ratio is 4 instead of 16 to preserve a high time resolution.
We pre-train the audio encoder on AudioSet to enhance its ability to recognize sound events.
Since queries are mostly short phrases in AudioGrounding, we use a single randomly initialized word embedding layer with mean pooling as the text encoder.

CLAP is trained on the same dataset as WSTAG training. 
We follow the configuration in \cite{xu2022dcase}, where CNN14 in PANNs and $\text{BERT}_\text{MEDIUM}$ are adopted as the audio encoder and text encoder.
During WSTAG training, we use the embedding after the text projection layer as the phrase embedding, with a size of 1024.


\subsection{Hyper-parameters}
We extract 64-dimensional log mel-spectrograms (LMS) as the audio feature.
The short-time Fourier transformation (STFT) window size and window shift are 32 ms and 10 ms.
The hidden size of the BiGRU is 256.
Dimensions of the audio and text embeddings are both 512.
The model is trained for at most 100 epochs with an early stop patience of 10 epochs and a batch size of 32, using Adam optimization.
The learning rate is initially 0.001 and reduced to $\frac{1}{10}$ if the validation loss does not improve for 3 epochs.

\subsection{Evaluation Metrics}

\paragraph*{\textbf{PSDS}}
Following previous practices~\cite{xu2021text,tang2021query}, polyphonic sound detection score (PSDS)~\cite{bilen2020framework} is used for evaluation.
It is the area under the PSD-ROC curve, which measures the relationship between the true positive rate (TPR) and the false positive rate (FPR).
Since the ground truth class of each phrase is unavailable, cross trigger is not considered here. 
We set $\rho_\text{DTC} = \rho_\text{GTC} = 0.5$, meaning that an output segment will be determined as correct if its overlap ratio with the ground truth segment is over 0.5.
However, $e_{max}$ is changed from 100 to 800, which is the maximum FPR value when calculating PSDS.
In SED, FPR is calculated as the average value of each class but in TAG FPR values of all classes are summed up since the ground truth class is unavailable. 
Therefore the maximum tolerant FPR should be increased.
PSDS is calculated using the toolkit from \cite{ebbers2022threshold}.

\begin{figure}
    \centering
    \includegraphics[width=0.9\linewidth]{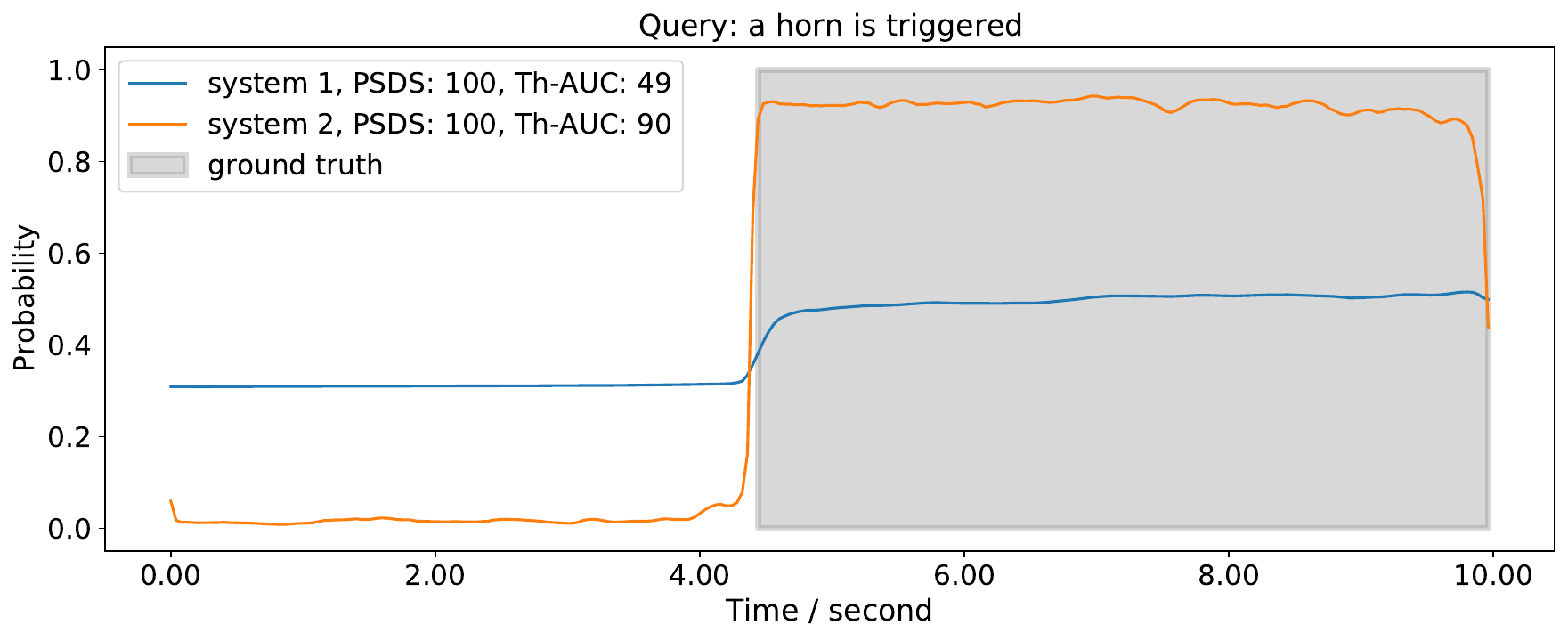}
    \caption{A comparison of evaluation on a sample using PSDS and Th-AUC. PSDS measures the performance under the best threshold while Th-AUC measures the performance over all possible thresholds.}
    \label{fig:system_compare_th_auc}
\end{figure}


\paragraph*{\textbf{Th-AUC}}
In PSDS evaluation, the TPR-FPR curve is plotted by calculating detection metrics under different thresholds.
However, we find that it does not measure the model's robustness against different thresholds since it considers only the optimal threshold for a specific FPR value to ensure the monotonicity of the PSD-ROC curve~\cite{bilen2020framework}.
For example, in \Cref{fig:system_compare_th_auc} two systems can both perform well under appropriate thresholds so their PSDS scores are the same.
However, system 2 performs better than system 1 under most thresholds.
To measure the model's robustness under a large scope of thresholds, we propose the area under $\mathrm{F}_1$-threshold curve (Th-AUC).
It measures the average performance under all possible thresholds.
A significant gap between the performance of two systems in \Cref{fig:system_compare_th_auc} is shown in terms of Th-AUC.

For both \textit{PSDS} and \textit{Th-AUC}, two results are reported.
One is calculated on the whole test set while the other is on a short-duration subset.
The subset is built by selecting segments that last for less than half of the whole audio clip for a more stringent evaluation.
Since many events last for the whole audio clip, models predicting high probabilities regardless of inputs still achieve high PSDS scores on the whole test set. 
On the short-duration subset, such models will get low scores.
Therefore, the subset can validate models' performance better.

\section{Results}
\label{sec:results}
In this section, we report our experimental results.
We provide an extensive analysis of the influencing factors and robustness of the proposed WSTAG approaches. 
Finally, a few qualitative results are given to show the performance and possible limits of our method.

\subsection{Influence of Pooling Strategies on Sentence-level WSTAG}

\begin{table}[ht]
    \centering
    \caption{WSTAG results using different pooling strategies.}
    \begin{tabular}{c|c|c|c|c|c}
    \toprule
     & \multirow{2}{*}{Pooling Strategy} & \multicolumn{2}{c|}{Whole} & \multicolumn{2}{c}{Short} \\
    \cline{3-6}
      &  & PSDS & Th-AUC & PSDS & Th-AUC \\
    \midrule
    & Mean & 29.9 & 38.0 & 5.5 & 3.7 \\
    Audio & Max & \textbf{42.7} & \textbf{48.9} & \textbf{33.8} & \textbf{42.4} \\
    Pooling & Linear Softmax & 32.4 & 40.4 & 6.2 & 8.7 \\
    & Exp. Softmax & 30.0 & 38.4 & 5.9 & 4.3 \\
    \midrule
    Text  & Mean & \textbf{42.7} & \textbf{48.9} & \textbf{33.8} & \textbf{42.4} \\
    Pooling & Sum & \textbf{42.7} & 43.9 & 33.4 & 38.1 \\
    \bottomrule
    \end{tabular}
    \label{tab:pooling_effect}
\end{table}

First, like the previous investigation on pooling strategies~\cite{xu2023investigating}, we analyze the influence of pooling strategies on sentence-level WSTAG.
Results are presented in the upper part of \Cref{tab:pooling_effect}.
We keep mean pooling for text the same.
The conclusion for the new framework aligns with that in the previous work that max pooling shows a significant advantage over other strategies.
As analyzed in \cite{xu2023investigating}, only one frame has a non-zero gradient using max pooling so it performs better under the weakly-supervised setting where gradients are likely to be misleading.
Then, the effect of text pooling is analyzed.
In contrast with the finding that sum pooling performs better in \cite{xu2023investigating}, mean pooling achieves superior performance in the new framework.
This may be attributed to the disparities in the granularity of similarity calculation in the two frameworks.
The old framework calculates frame-word similarities so functional words affect the aggregated similarity to a large extent.
However, the new framework directly calculates frame-phrase similarities so the model may be trained to ignore functional words when encoding phrases.
As a result, the negative effect of mean pooling is alleviated.
Besides, similar PSDS scores but the gap in Th-AUC scores show that the two pooling strategies perform similarly under the optimal threshold but mean pooling performs better in a larger scope of thresholds.
This indicates that in sum pooling, models are trained to assign lower scores to positive phrase queries to reduce the score of negative captions with both positive and negative phrases, leading to their worse robustness against thresholds.

\begin{table*}[ht]
    \centering
    \caption{Results of different WSTAG approaches. The best WSTAG and SSTAG results are highlighted in bold. For all WSTAG methods, we use the same model architecture.}
    \begin{tabular}{c|l|c|c|c|c}
    \toprule
     & \multirow{2}{*}{Approach} & \multicolumn{2}{c|}{Whole} & \multicolumn{2}{c}{Short} \\
    \cline{3-6}
    & & PSDS & Th-AUC & PSDS & Th-AUC \\ 
    \midrule
    \multicolumn{6}{c}{\cellcolor[HTML]{f2f2f2}
    \textit{Our Proposed WSTAG}}\\
    \midrule
    \multirow{2}{*}{Sentence-level WSTAG (Ours)} & A-Mean + T-Mean & 29.9 & 38.0 & 5.5 & 3.7 \\
     & A-Max + T-Mean & 42.7 & 48.9 & 33.8 & 42.4 \\
    \midrule
    \multirow{6}{*}{Phrase-level WSTAG (Ours)} & Random Sampling & 43.7 & 46.5 & 34.5 & 43.2 \\
    & \hspace{2em} + Self-Supervision & 48.7 & 50.5 & 40.1 & 46.7 \\
    \cline{2-6}
     & Similarity-based Sampling & 52.6 & 53.9 & 43.7 & 48.5 \\
     & \hspace{2em} + Self-Supervision & 55.7 & \textbf{57.1} & 46.2 & 50.8 \\
     \cline{2-6}
     & Clustering-based Sampling & 52.9 & 54.2 & 44.4 & 49.3 \\
     & \hspace{2em} + Self-Supervision & \textbf{56.5} & \textbf{57.1} & \textbf{47.6} & \textbf{51.7} \\
     \midrule
    \multicolumn{6}{c}{\cellcolor[HTML]{f2f2f2}
    \textit{Baseline TAG Methods (from Previous Works or Adaptation)}}\\
    \midrule
    Previous WSTAG Methods & CDur-word Alignment~\cite{xie2022unsupervised} & 35.2 & 35.4 & 6.8 & 5.3 \\
    \midrule
    \multirow{6}{*}{Previous SSTAG Methods} & CDur-w2vmean~\cite{xu2021text} & 56.0 & 40.2 & 39.8 & 23.0 \\
     & CDur-QGCA~\cite{tang2021query} & 57.4 & 47.2 & 43.2 & 31.5 \\
     & Conditional CDur (Adapted from \cite{kong2020source}) & 57.6 & 48.8 & 43.1 & 35.7 \\
     & CRNN-w2vmean & 58.3 & 42.3 & 43.3 & 25.6 \\
     & CRNN-BERT & 60.4 & \textbf{56.0} & 48.5 & \textbf{44.3} \\
     & CRNN-QGCA & \textbf{62.8} & 53.1 & \textbf{49.8} & 38.4\\
    \bottomrule
    \end{tabular}
    \label{tab:w2vmean_approach_compare}
\end{table*}

\subsection{Comparison between Approaches}

In the upper half of \Cref{tab:w2vmean_approach_compare}, we present the performance of our proposed WSTAG approaches.
We use the same architecture for all approaches.
Compared with the baseline sentence-level WSTAG using mean pooling for both audio and text, significant improvement is achieved by replacing audio mean pooling with max pooling.
For phrase-level WSTAG, we use a phrase number $n$ (see \Cref{alg:clustering_sampling}) of 32 in all experiments.
The similarity threshold $\tau$ in similarity-based sampling is 0.5.
Clustering-based sampling uses k-means clustering with a cluster number of 32.
All models use linear softmax pooling.
Results show that phrase-level WSTAG is superior to sentence-level approaches.
Even the most straightforward random sampling can achieve comparable performance to the best-performing sentence-level WSTAG.
With the proposed advanced sampling strategies based on similarity and clustering, significant improvement is achieved.
An absolute 10\% improvement in terms of $\text{PSDS}_\text{whole}$ is achieved, indicating that our proposed sampling strategies significantly enhance the quality of training labels.
Clustering-based sampling shows an advantage over similarity-based sampling.
This validates the benefit of a more balanced distribution of events corresponding to negative phrases.

Regardless of the sampling strategy, the involvement of self-supervision brings substantial performance improvement.
The refinement of clip-level labels and the availability of frame-level labels from a pre-trained model further improves the label quality.
In particular, the introduction of frame-level pseudo labels enables the model to receive frame-level supervision without ground truth frame-level labels, significantly enhancing the temporal localization accuracy.


\subsection{Comparison with Previous Methods}


We list several previous TAG methods for comparison in the lower half of \Cref{tab:w2vmean_approach_compare}.
Since there are few works on TAG, we include methods adapted from related works for comparison.
Conditional CDur uses a method similar to \cite{kong2020source} to fuse text into the SED model.
The text feature is added to each output channel of feature maps to predict queries.
CRNN-w2vmean and CRNN-QGCA are adapted from \cite{xu2021text} and \cite{tang2021query} by replacing the CDur audio encoder with the CRNN in this work.
CRNN-BERT further improves CRNN-w2vmean by using a frozen $\text{BERT}_\text{BASE}$ text encoder.

These adapted methods provide much stronger baselines.
Our proposed WSTAG model achieves competitive performance in terms of PSDS while outperforming all SSTAG methods in Th-AUC.
WSTAG automatically learns event patterns from weak labels and leverages a large amount of audio-text data to avoid the drawback of imprecise timestamp annotations of short events.
Therefore, WSTAG demonstrates a greater advantage in short-duration event detection.
Additionally, the imprecision of timestamp annotations around event boundaries may cause SSTAG models to produce outputs that are more inclined towards moderate values, thus lacking robustness to threshold variations.
In contrast, WSTAG mitigates this issue, achieving favorable results across a wider range of thresholds and thereby attaining a higher Th-AUC.



\subsection{Evaluation on SED}

\begin{table}[htpb]
    \centering
    \caption{SED performance in terms of F$_1$ score on the DESED test set. We only list the average score and scores of some classes due to limited space.}
    \begin{tabular}{c|ccccc|c}
    \toprule
    Model & Speech & Dog & Cat & Water & Frying & Average \\
    \midrule
    PANNs~\cite{kong2020panns} & 69.7 & 35.8 & 36.3 & 30.6 & 9.3 & 35.1\\
    HTS-AT~\cite{chen2022hts} & 46.8 & 48.0 & 67.7 & \textbf{43.0} & \textbf{60.3} & 48.4 \\
    WSTAG (Ours) & \textbf{84.5} & \textbf{80.8} & \textbf{84.2} & 34.5 & 54.6 & \textbf{58.0} \\
    \bottomrule
    \end{tabular}
    \label{tab:desed_result}
\end{table}

To assess the generalizability of our model, we further evaluate our model on the DESED test set~\cite{serizel2020sound}, a sound event detection dataset.
Following \cite{chen2022hts}, we calculate F$_1$ scores under PSDS settings, as Th-AUC does.
We compare our model with PANNs~\cite{kong2020panns} and HTS-AT~\cite{chen2022hts} as they are not trained directly on DESED either.
Our model is trained on AudioCaps, which is only $\frac{1}{20}$ of AudioSet, where PANNs and HTS-AT are trained.
The result in \Cref{tab:desed_result} shows that our method efficiently leverages text annotation of audio clips, outperforming PANNs and HTS-AT significantly with much smaller training data.
For common sounds like speech and dog, our model performs well on DESED without being exposed to its training set, validating the robustness of our model.

\subsection{Analysis on Influencing Factors of Phrase-level WSTAG}

In this part, we extensively analyze the influence of several factors on phrase-level WSTAG.
When we analyze one factor, we keep all other settings the same as that in the previous part.
For simplicity, we do not involve self-supervision since it requires two training stages and we only list the results of PSDS scores on the short-duration subset ($PSDS_{short}$).

\subsubsection{Data Size}

\begin{figure}[ht]
    \centering
    \includegraphics[width=0.9\linewidth]{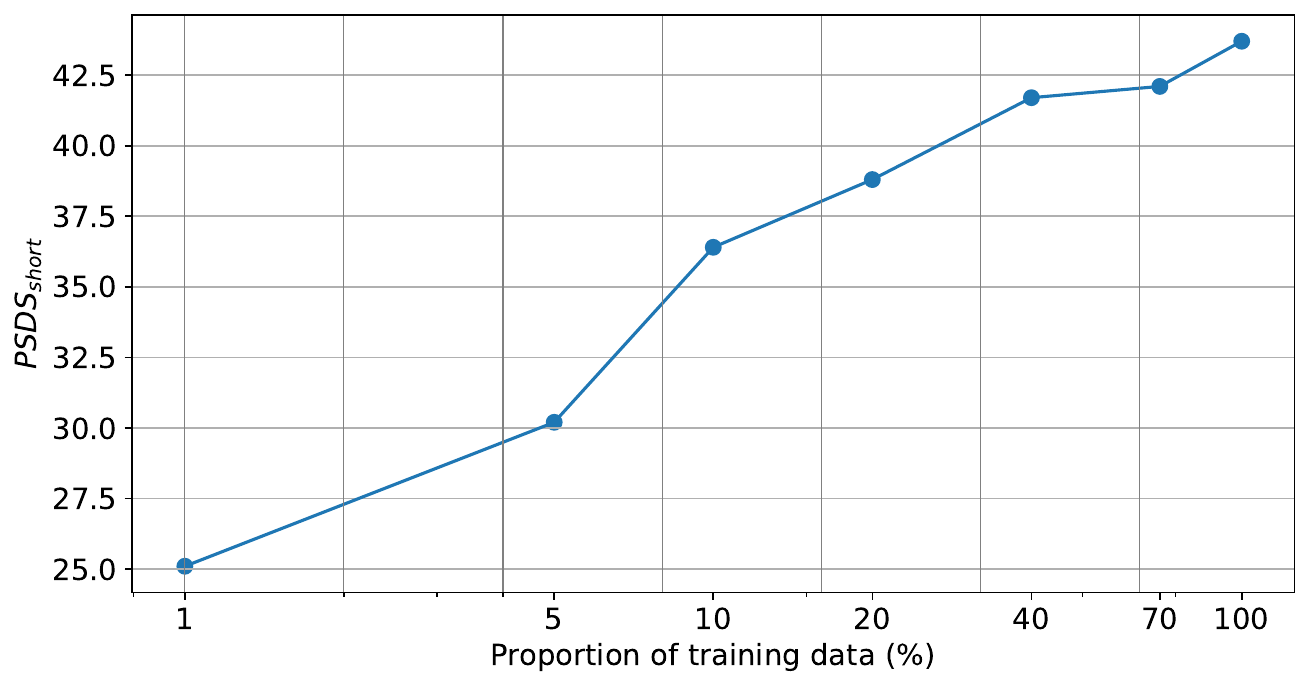}
    \caption{Analysis of WSTAG using different amounts of training data.}
    \label{fig:data_size_analysis}
\end{figure}

Since the advantage of WSTAG is to leverage large-scale weakly-annotated audio-text data for training, we analyze the influence of data size on the performance.
For simplicity, the experiments are based similarity-based sampling strategy so the training of clustering models using different portion of data is not needed. 
in \Cref{fig:data_size_analysis}.
Generally, WSTAG performance improves with an increase in the dataset size.
When the dataset is relatively small, such as 5\% of the total, the performance gains from increasing the dataset size are quite pronounced.
However, as the dataset size grows to around 50\% (25K pairs), the performance improvement becomes limited.
Since the dataset size is already much larger than the strongly-supervised one (5K pairs) with most categories covered, further increasing the size may have a limited impact on frequent categories.
In addition, the improvement in infrequent categories contributes relatively little to the overall performance, resulting in limited overall gains.

\subsubsection{Pooling Strategies}

\begin{figure}[ht]
    \centering    \includegraphics[width=0.8\linewidth]{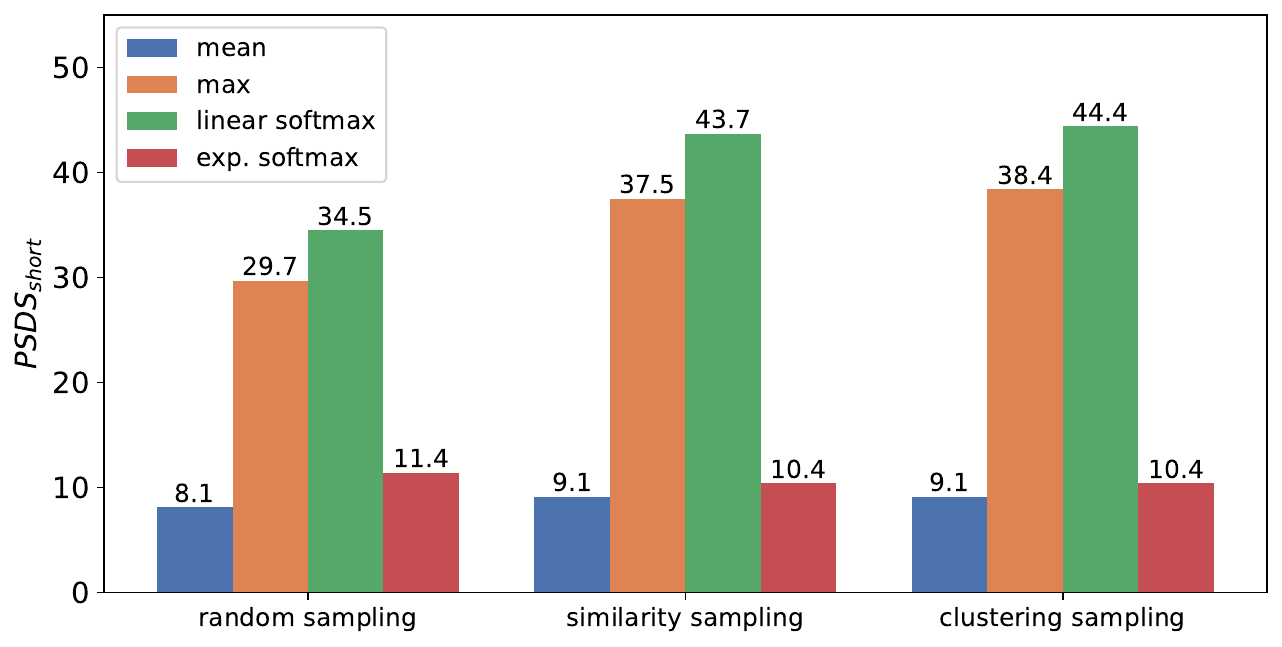}
    \caption{Comparison of different pooling strategies.}
    \label{fig:pooling_analysis}
\end{figure}

\Cref{fig:pooling_analysis} shows the performance of phrase-level WSTAG approaches using different pooling strategies.
Same as audio pooling in sentence-level WSTAG, we compare strategies in \Cref{subsec:pooling}.
The result is similar to sentence-level WSTAG in that mean pooling performs the worst.
The violation of the SMI assumption dramatically hurts the performance.
However, linear softmax consistently outperforms other strategies in phrase-level WSTAG.
Max pooling still achieves good performance but lags behind linear softmax to a large extent.
This comparison between sentence- and phrase-level WSTAG results suggests that pooling strategies heavily rely on suitable loss functions to work well: linear softmax is compatible with BCE loss while max pooling performs well under the contrastive loss.

\subsubsection{Phrase Numbers}

\begin{figure}[ht]
    \centering
    \includegraphics[width=0.8\linewidth]{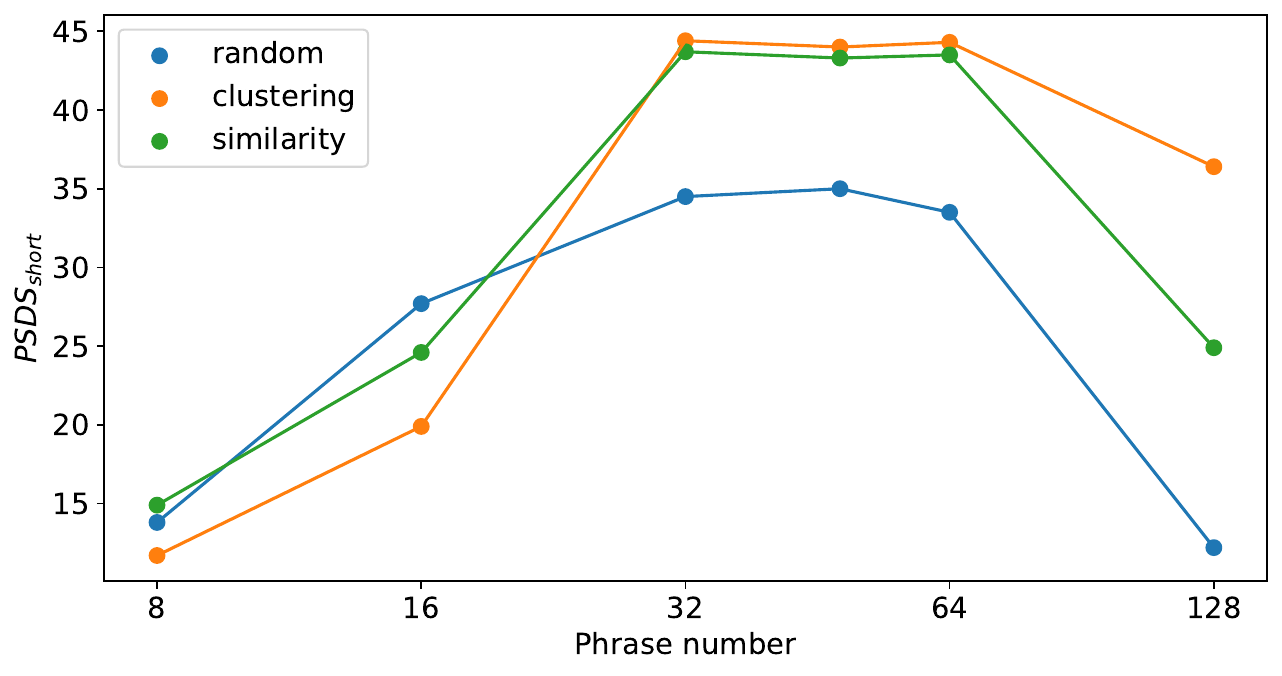}
    \caption{Comparison of different phrase numbers.}
    \label{fig:phrase_number_analysis}
\end{figure}

We analyze the influence of using different phrase numbers $n$.
More negative phrases are sampled as $n$ increases.
Results are shown in \Cref{fig:phrase_number_analysis}.
For either sampling strategy, the phrase number imposes a significant impact on the performance and the optimal phrase number is around 32 and 64.
If $n$ is too small, phrases describing various sound events are not sampled during training.
However, when $n$ is too large, the probability that false negative phrases are sampled also becomes larger.
It is observed that as $n$ increases to 128, the advantage of the advanced sampling strategies over random sampling grows bigger.
However, when $n$ is much lower than the optimal value, the clustering-based sampling strategy leads to the worst result.
This may be caused by the distribution of events.
As $n$ is much lower than the total sound event number, clustering-based sampling can only sample phrases covering a part of sound events.
Even the most frequent events cannot be detected well.
However, random sampling has more access to these frequent events.
Therefore, random sampling achieves better performance when $n$ is small.

\subsubsection{Phrase Embeddings}


\begin{table}[h]
    \centering
    \caption{Comparison of different phrase embeddings.}
    \begin{tabular}{c|cc}
    \toprule
    Sampling Strategy & Similarity Sampling & Clustering Sampling \\
    \midrule
    $\text{SBERT}_\text{roberta-large}$ & 42.2 & 28.3 \\
    $\text{CLAP}_\text{bert-medium}$ & 43.7 & 44.4 \\
    \bottomrule
    \end{tabular}
    \label{tab:text_embedding_analysis}
\end{table}

Here we analyze the influence of different phrase embeddings on negative sampling strategies in \Cref{tab:text_embedding_analysis}.
We compare a pure semantic encoder, Sentence-BERT~\cite{reimers2019sentence} with the audio-centric CLAP.
We use $\text{Roberta}_\text{LARGE}$ as its embedding size is the same as CLAP.
Without audio-text learning, Sentence-BERT gives the similarity between phrases solely based on the semantic meaning.
Therefore, CLAP consistently achieves better results than Sentence-BERT.
Especially in clustering-based sampling, phrases from clusters of positive phrases are never sampled.
The reliability of clustering is crucial to the result so better text embeddings can improve the performance significantly.

\subsubsection{Cluster Numbers}

\begin{figure}[ht]
    \centering
    \includegraphics[width=0.8\linewidth]{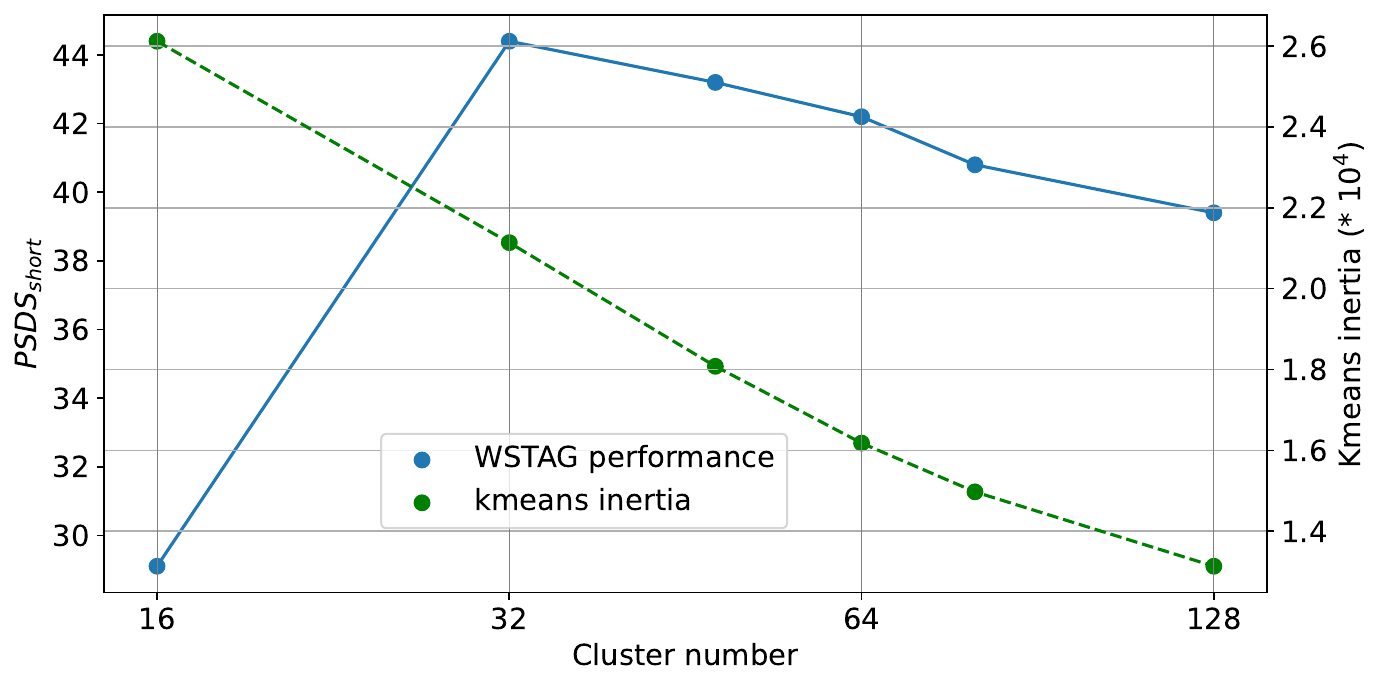}
    \caption{Comparison of different cluster numbers in clustering-based sampling.}
    \label{fig:cluster_number_analysis}
\end{figure}

The number of cluster centers $n_c$ is an important factor in clustering.
We present the results of clustering-based sampling using different $n_c$ in \Cref{fig:cluster_number_analysis}.
Here the k-means clustering algorithm is exclusively used.
We also plot the clustering inertia for reference.
It measures the distance of samples to their closest cluster center.
Although the inertia keeps decreasing as $n_c$ grows, WSTAG performance starts declining after $n_c$ reaches 32.
Too small or too large $n_c$ hurt the performance.
When $n_c$ is too small, phrases describing the same general sound but belonging to different fine-grained sounds are grouped into the same cluster (called false-merging).
For example, phrases describing all animals are grouped into the same cluster when $n_c$ is extremely small.
During evaluation, there will be many false positive predictions for fine-grained sound event, e.g., dog barking or pig oinking.
Therefore, the performance declines dramatically with only 16 clusters.
In contrast, a large cluster number will force phrases describing the same sound event to group into different clusters (we call it false-splitting), especially for sound events with a large number of phrases like male speech.
The model will thus estimate lower probabilities for these events when they are present.
It can be observed from the comparison of performance under small and large $n_c$ that the false-merging problem is more harmful than false-splitting.

\subsubsection{Clustering Algorithms}

\begin{table}[h]
    \centering
    \caption{Comparison of different cluster methods in clustering-based sampling.}
    \begin{tabular}{c|ccc}
    \toprule
    Clustering Method & Kmeans & Spectral & Agglomerative  \\
    \midrule
    $\text{PSDS}_\text{short}$ & 44.4 & 41.5 & 38.2 \\
    \bottomrule
    \end{tabular}
    \label{tab:cluster_method_analysis}
\end{table}

Since the clustering result is critical to clustering-based sampling, we analyze the influence of different clustering algorithms.
The result is shown in \Cref{tab:cluster_method_analysis}.
Besides k-means clustering, we also investigate spectral clustering and agglomerative clustering.
K-means performs the best.
It may be attributed to the fact that k-means is a general clustering algorithm but the other two are more suitable for specific data types.
Compared with phrase embeddings and cluster numbers, our approach is not sensitive to the clustering algorithm.

\subsubsection{Similarity Thresholds}

\begin{figure}[ht]
    \centering
    \includegraphics[width=0.95\linewidth]{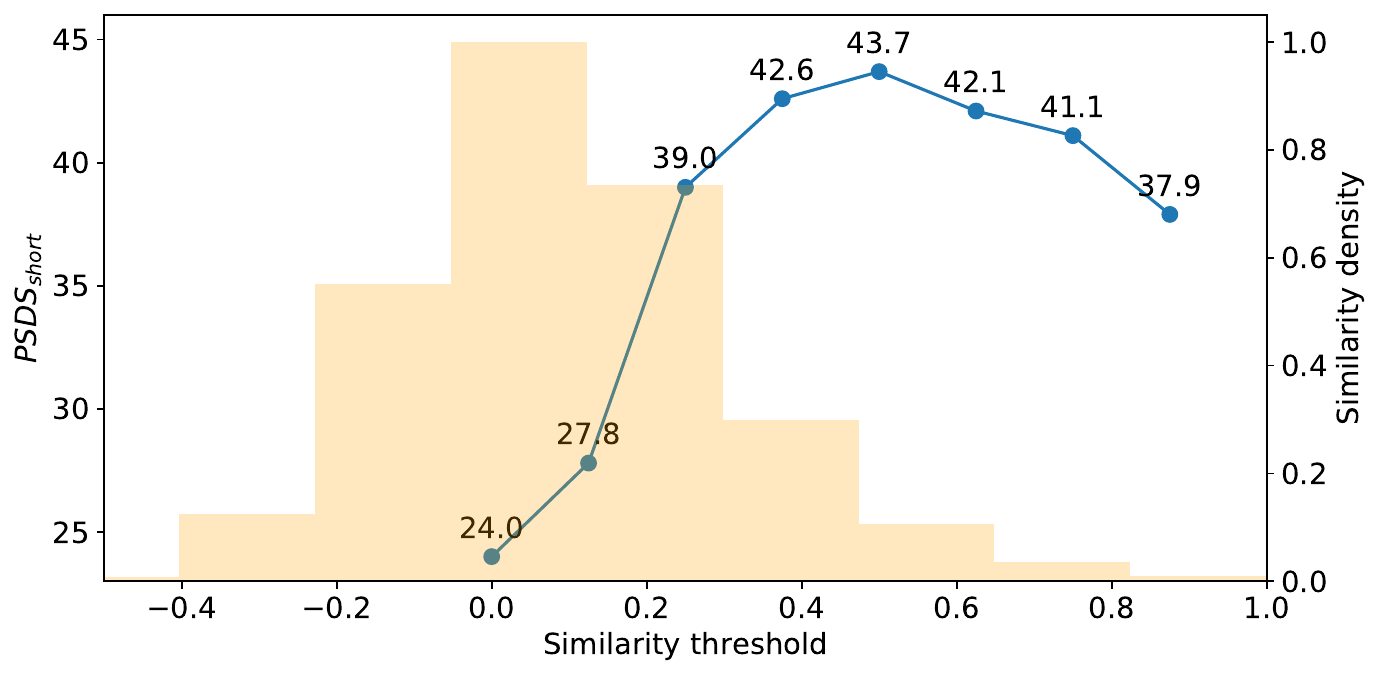}
    \caption{Comparison of different similarity thresholds $\tau$ in similarity-based sampling.}
    \label{fig:sim_threshold_analysis}
\end{figure}

\begin{figure}[ht]
    \centering
    \includegraphics[width=0.75\linewidth]{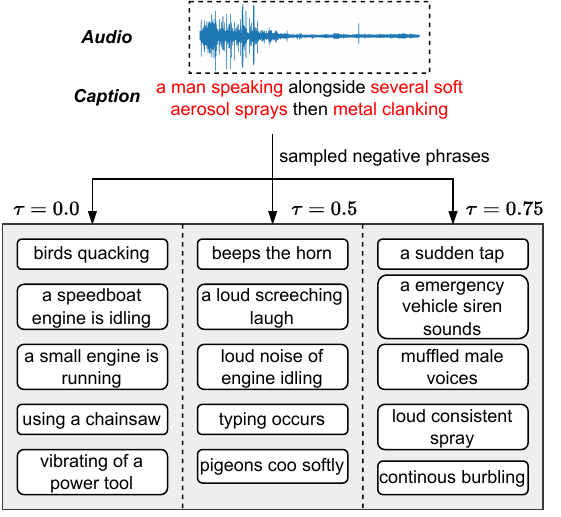}
    \caption{An example of negative phrases sampled under different similarity thresholds $\tau$.}
    \label{fig:sim_sampling_egs}
\end{figure}

For similarity-based sampling, an important hyper-parameter is the similarity $\tau$ used to determine whether a phrase is negative for an audio clip.
Its influence is investigated and presented in \Cref{fig:sim_threshold_analysis}.
The distribution of similarities between different phrases is also plotted in terms of the normalized density.
$\tau = 0.5$ achieves the best result.
When $\tau$ becomes larger, the algorithm fails to filter out many false negative phrases.
When $\tau$ is not large enough, there is a risk of ignoring hard negative phrases (e.g., ignoring negative phrases like \textit{``another man speaking''} in a recording of a single man speaking).
However, the inclusion of false negative phrases (e.g., taking \textit{``male voice ''} as negative) has a larger influence than ignoring hard negative phrases, as indicated by the drop in the performance when $\tau$ keeps increasing.
The discrimination of hard negative phrases may need a better CLAP.

If $\tau$ becomes too small, the number of candidate phrases also decreases a lot.
Although there are still a large number of phrase pairs with negative similarities, phrases corresponding to many sound events are never sampled under $\tau = 0$, resulting in poor performance though the sampled phrases are definitely negative.
An example is shown in \Cref{fig:sim_sampling_egs}.
When $\tau = 0.0$, the diversity of sampled phrases is limited: 4 of 5 randomly sampled phrases describe the humming or vibration of engines.
When $\tau$ increases to 0.75, sampled phrases become much more diverse but false negative phrases are also sampled, e.g., \textit{``loud consistent spray''}.
$\tau = 0.75$ performs much better than $\tau = 0.0$, suggesting that the diversity of sampled phrases and corresponding events is more important than the accuracy of phrase labels.

\begin{figure}[ht]
    \centering
    \includegraphics[width=0.95\linewidth]{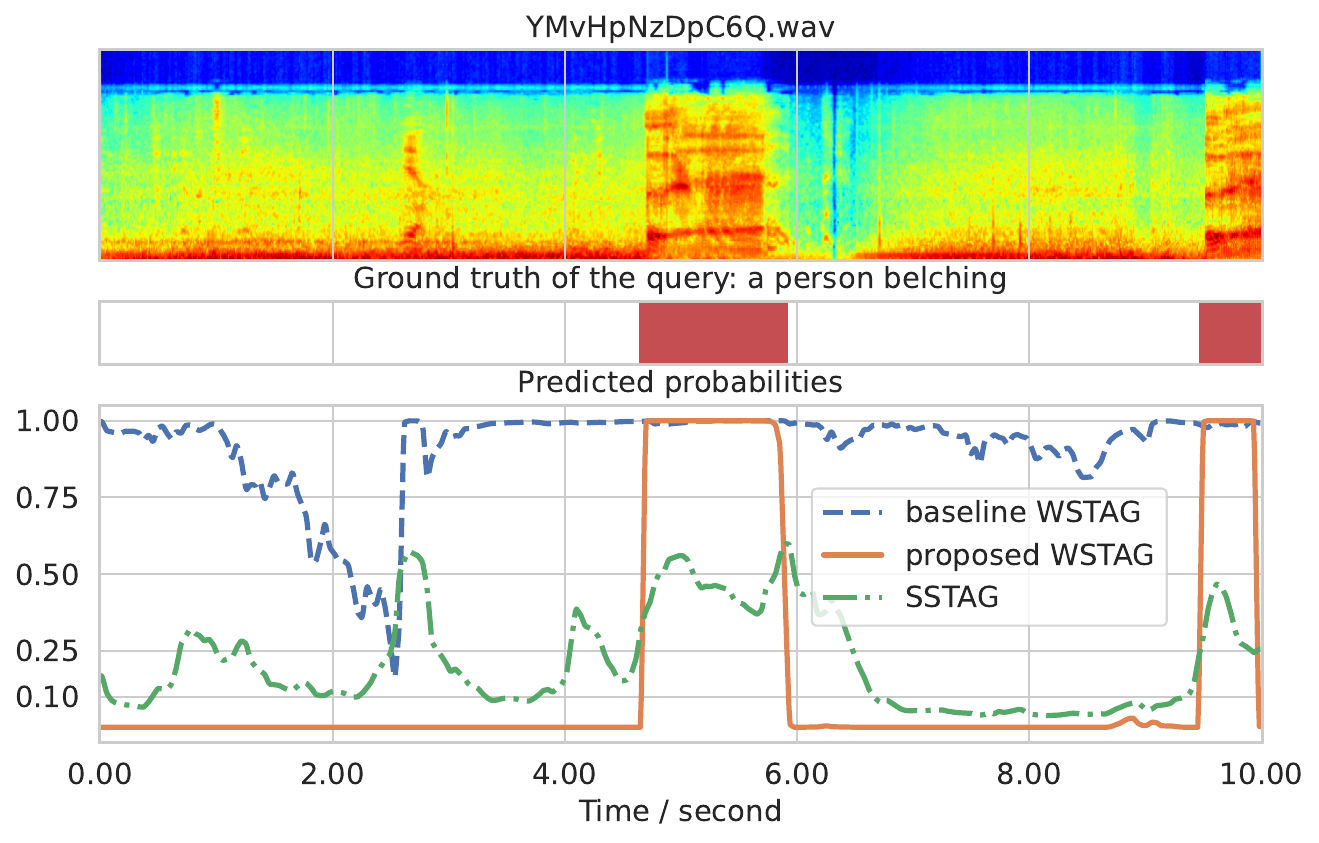}
    \caption{An audio-phrase pair sample comparison on (1) the baseline WSTAG system; (2) our proposed WSTAG system; and (3) SSTAG system. Best viewed in color.}
    \label{fig:system_compare_visualize}
\end{figure}

\begin{figure*}[ht]
    \centering
    \includegraphics[width=0.9\linewidth]{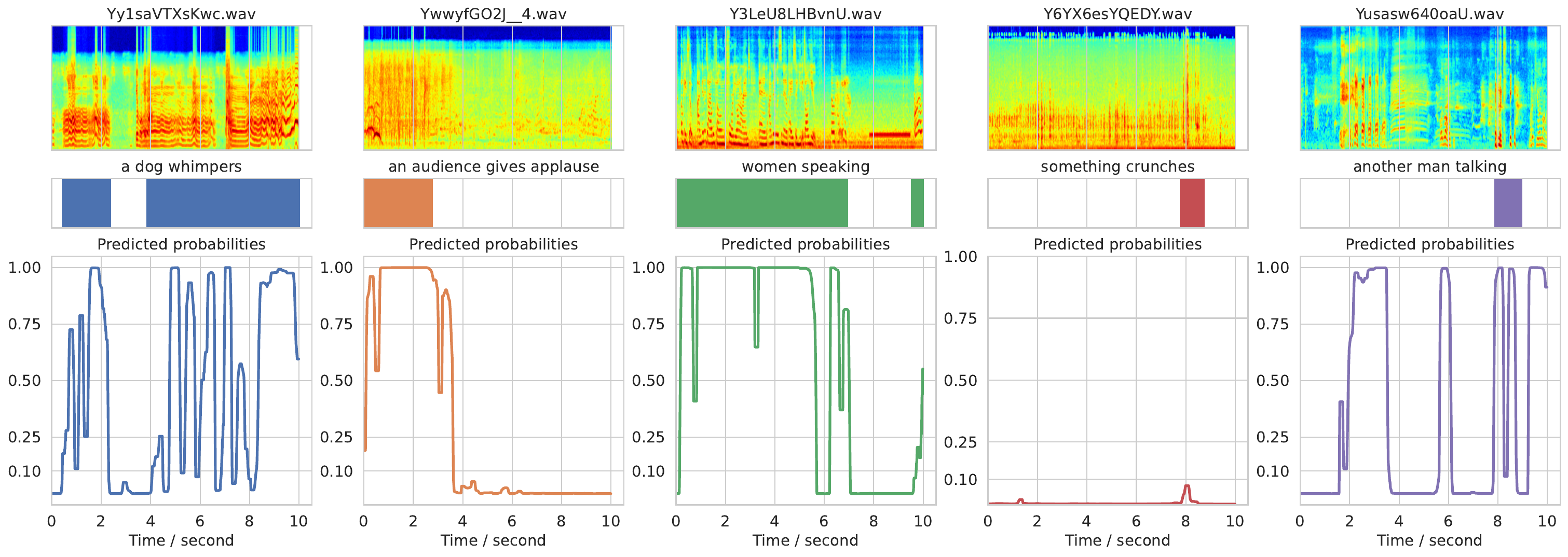}
    \caption{Predictions of our proposed WSTAG system for five samples, including 3 successful cases and 2 failure cases.}
    \label{fig:proposed_examples}
\end{figure*}

\subsection{Qualitative Results}

In this part, we provide an analysis of our proposed WSTAG system through visualization.
We first compare our best-performing WSTAG system with the baseline WSTAG and SSTAG systems.
The baseline WSTAG system is the sentence-level one with mean pooling while our proposed WSTAG system is the phrase-level one with clustering-based sampling and self-supervision.
Then we provide several qualitative data samples, including both successful and failure cases, to show the model's strengths and limitations.
Finally, we compare our WSTAG system with an SED system to highlight the advantage of TAG over SED.


We randomly sample an audio-phrase pair from the short-duration subset and the comparison of different approaches is shown in \Cref{fig:system_compare_visualize}.
The phrase query \textit{``a person belching''} corresponds to the class \textit{``Hiccup''} in AudioSet, which is an infrequent class with only 931 samples in the total 2.1 M audio clips.
The baseline system is sentence-level WSTAG with mean pooling for both audio and text.
It fails to detect the segments since it assigns high probabilities to all non-silent segments, regardless of the sound type.
The violation of SMI assumption results in its bad performance.
Although trained on strongly-annotated data, SSTAG is incapable of producing perfect predictions neither.
This may be attributed to the scarcity of the event \textit{``Hiccup''} in the strongly-annotated dataset.
In contrast, our proposed phrase-level WSTAG system accurately predicts the onsets and offsets.
Since the acoustic characteristics of hiccups are notable, WSTAG systems are trained to recognize such sounds from a variety of audio clips.

Furthermore, several examples of our proposed system's predictions are shown in \Cref{fig:proposed_examples}.
The first three are successful cases, where accurate predictions can be obtained by using a standard threshold of 0.5 and post-processing like median filtering.
As WSTAG methods do not involve timestamp labels for training, the model learns sound patterns from the comparison between recordings with and without specific events.
No preference like merging short segments is introduced in this training paradigm. 
Therefore, for short and loud sounds, the probability predicted by the model oscillates sharply between high and low values, as shown in the first example.
The rest two examples are typical failure cases.
The query of the first example is \textit{``something crunches''}, which is often in the background with a low loudness.
It is difficult for the model to learn the characteristics of such background sounds with the interference of predominant sounds.
In the last example, the caption is a detailed description: \textit{``$\cdots$ and a man talking followed by another man talking''}, with the phrase query \textit{``another man talking''}.
It requires the model to recognize that there are two men speaking and that the desired output is the speech segment of the second man.
The model predicts all speech segments.
The behavior is attributed to the characteristics of training data: most captions just describe present sound events without details such as the speaker number or the temporal relationship between events.
Since AudioCaps only cover a limited set of frequent, daily non-musical sounds, it is challenging for our WSTAG to detect rare, unseen sounds accurately.
These failure cases indicate the limitations of our model in detecting background sounds and handling queries that seldom occur in the training set, including unseen sound categories and queries with details like the speaker identity.

\begin{figure}
    \centering
    \includegraphics[width=0.9\linewidth]{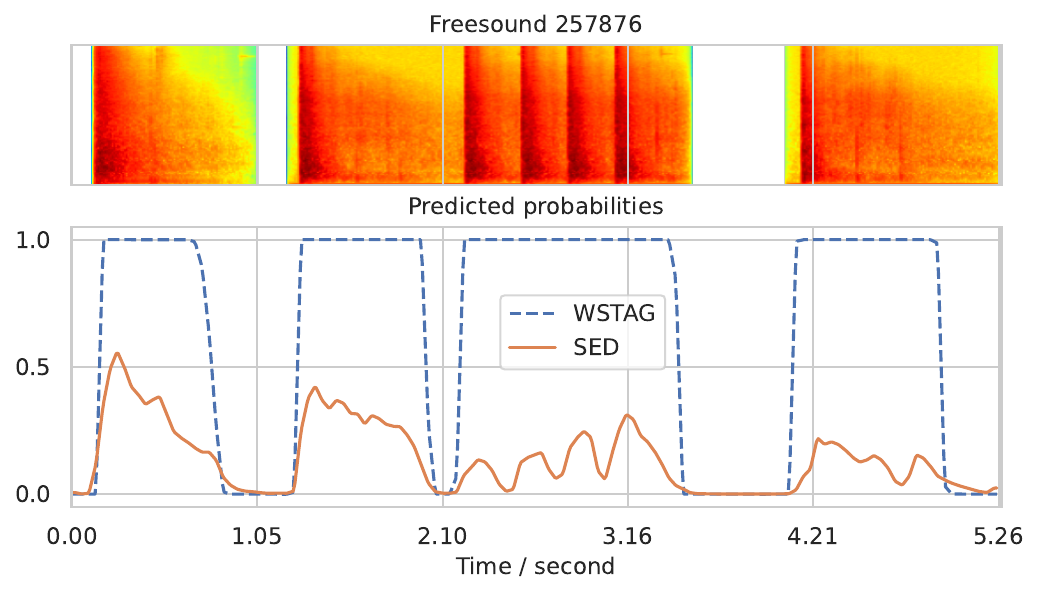}
    \caption{Comparison between WSTAG and SED models on a Freesound sample.}
    \label{fig:compare_sed_grounding}
\end{figure}

Finally, we present the advantage of TAG over SED models by comparing their predictions.
We choose a gunshot sample from out-of-distribution Freesound and the comparison is shown in \Cref{fig:compare_sed_grounding}.
The SED model is trained on strongly-annotated AudioSet subset~\cite{hershey2021benefit}, which is twice the size of AudioCaps.
In spite of the larger training data size, the SED model gives relatively low probabilities when the gunshot occurs.
However, the prediction of the WSTAG model is near optimal.
We speculate the problem of SED originated from AudioSet labels.
Due to the duplication problem of AudioSet labels~\cite{gong2021psla}, sound events such as gunshot are sometimes missing in annotations.
During training, probabilities predicted for these events in data containing them are discouraged, resulting in lower predicted probabilities when these events occur during inference.
In contrast, annotators tend not to overlook prominent foreground sound events when describing an audio clip, thus avoiding this problem.
Therefore, the advantage of natural language, compared with categorical systems, leads to TAG's advantage over SED.

\section{Conclusion}
\label{sec:conclusion}

In this paper, we explore TAG, the intersection of natural language processing and sound event detection, which aims to predict timestamps for sound events described by language.
A major challenge in this field is the scarcity of strongly-labeled data, underscoring the importance of WSTAG research.
With a lower requirement for annotations, WSTAG can make use of large-scale audio-text datasets for training.
We first analyze the limitations of sentence-level WSTAG pooling strategies and investigate different pooling strategies.
The best results are achieved with audio max pooling and text mean pooling.
We then propose phrase-level WSTAG to narrow the training/test gap in the textual modality.
Furthermore, we propose advanced negative sampling strategies and self-supervision to improve weak label accuracy and reduce the gap in the audio modality.
Our phrase-level WSTAG outperforms the SSTAG system with the same architecture and is close to the SOTA SSTAG system, with strong performance on short-duration sounds.
The evaluation on an unseen SED dataset validates the generalization ability of our model.
We also analyze several factors impacting phrase-level WSTAG and visualize the advantages and limitations of our approach.


%



\section*{Acknowledgment}
This work has been supported by Shanghai Municipal Science and Technology Major Project (2021SHZDZX0102) and Jiangsu Technology Project (No.BE2022059-2).
We would like to thank Mark D. Plumbley, who provided valuable insights and assistance in polishing this manuscript.

\ifCLASSOPTIONcaptionsoff
  \newpage
\fi



\bibliographystyle{IEEEtran}
\bibliography{refs}

\begin{IEEEbiography}[{\includegraphics[width=1in,height=1.25in,clip,keepaspectratio]{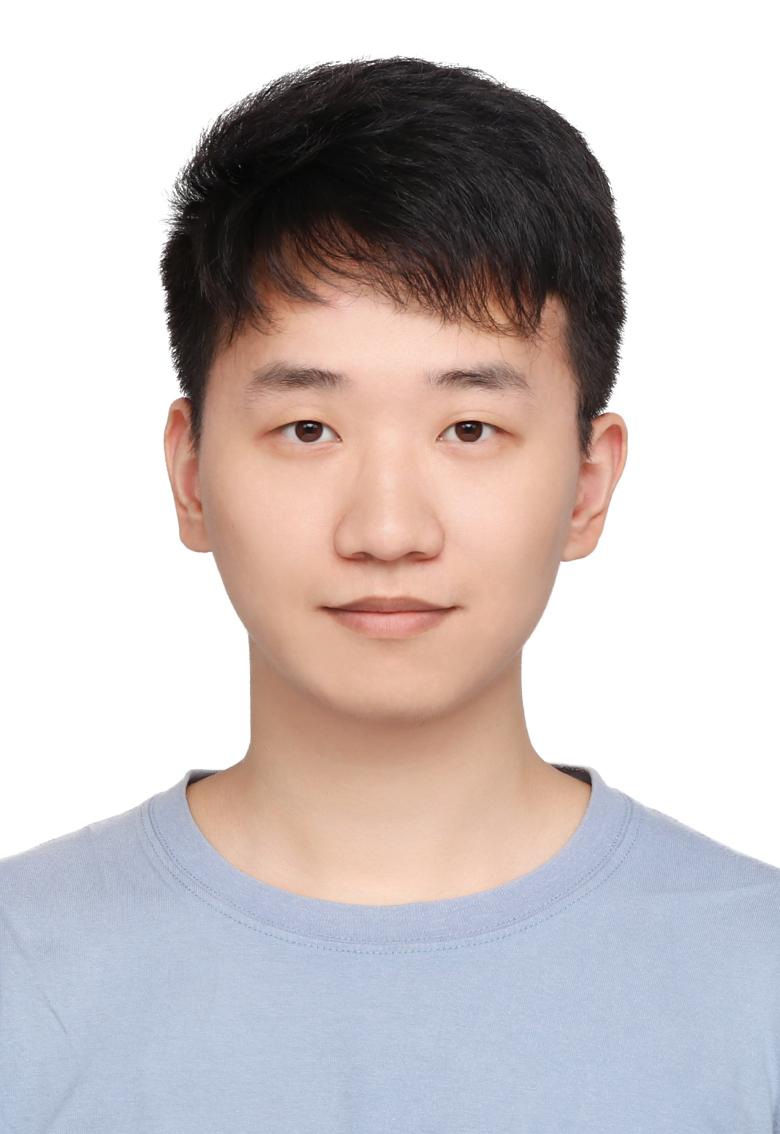}}]{Xuenan Xu}
received his B.S. degree from Shanghai Jiao Tong University in 2019. He is currently working towards his Ph.D. degree with the Department of Computer Science and Engineering, Shanghai Jiao Tong University. His main research interests include audio understanding and generation and multi-modal learning.
\end{IEEEbiography}
\begin{IEEEbiography}[{\includegraphics[width=1in,height=1.25in,clip,keepaspectratio]{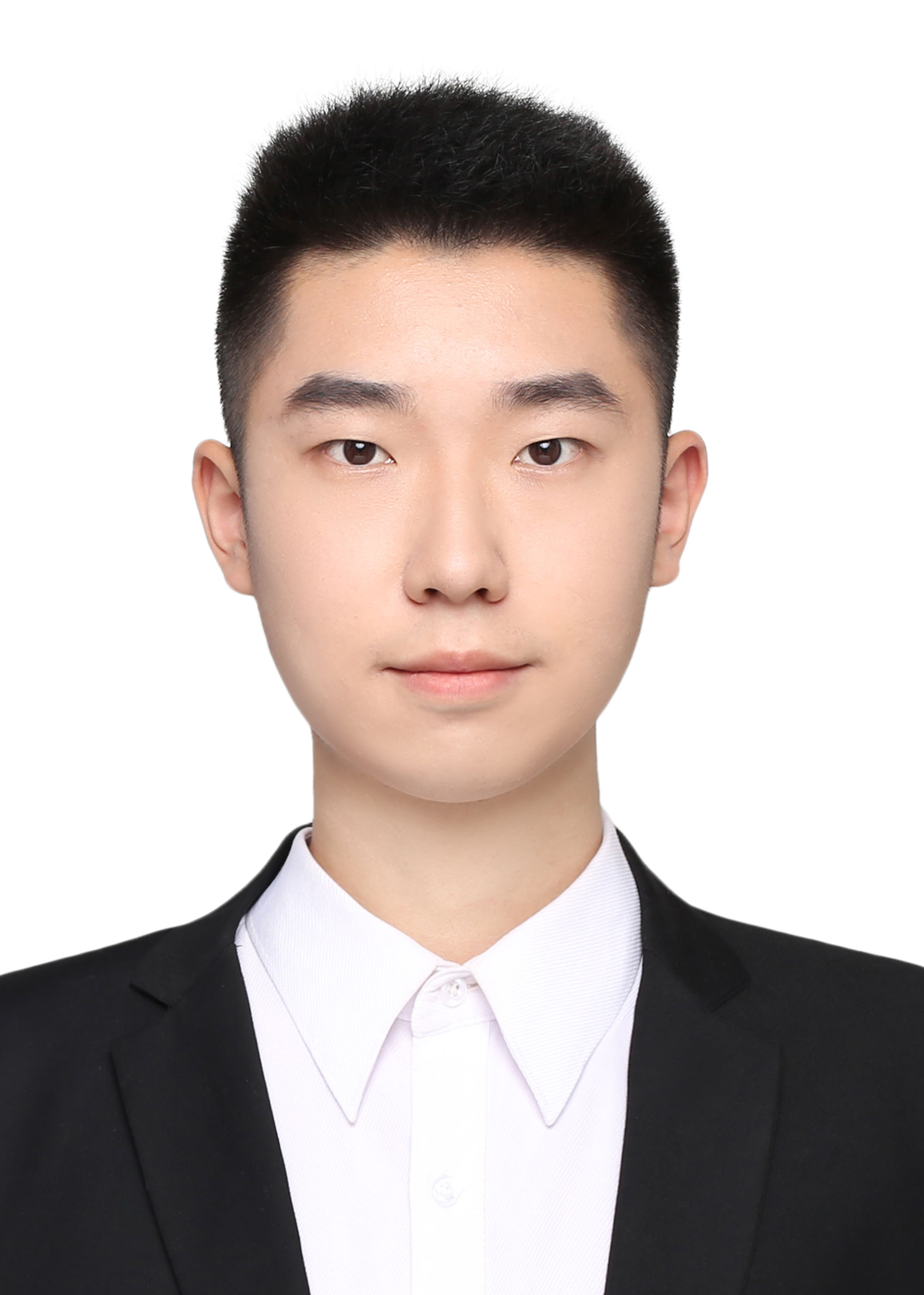}}]{Ziyang Ma}
received the B.Eng. degree in computer science from Shandong University in 2022. He is currently working toward the Ph.D. degree with the Department of Computer Science and Engineering, Shanghai Jiao Tong University. His research interests focus on speech, language, audio and music processing with Self-Supervised Learning (SSL) and Large Language Model (LLM).
\end{IEEEbiography}
\begin{IEEEbiography}[{\includegraphics[width=1in,height=1.25in,clip,keepaspectratio]{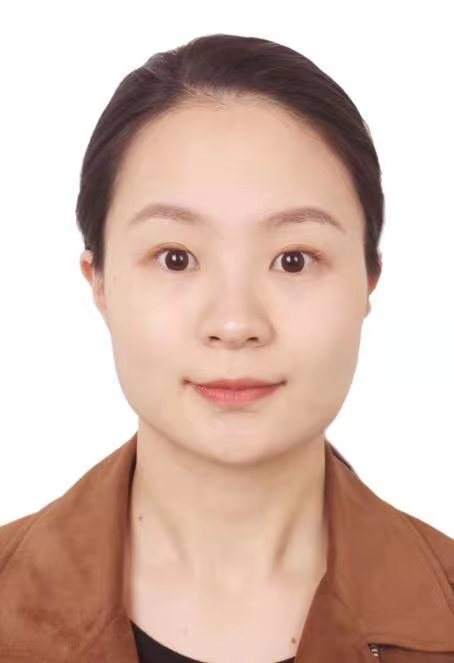}}]{Mengyue Wu}
received her B.S. and B.A. from Beijing Normal University in 2011 and was awarded Ph.D. from the University of Melbourne in 2017. She is currently an Assistant Professor in Computer Science and Engineering Department, Shanghai Jiao Tong University, China. Her main research interests lie in the area of audio- and language- based human machine interaction including audio processing, multimedia processing, and medical application of these technologies. 
\end{IEEEbiography}
\begin{IEEEbiography}[{\includegraphics[width=1in,height=1.25in,clip,keepaspectratio]{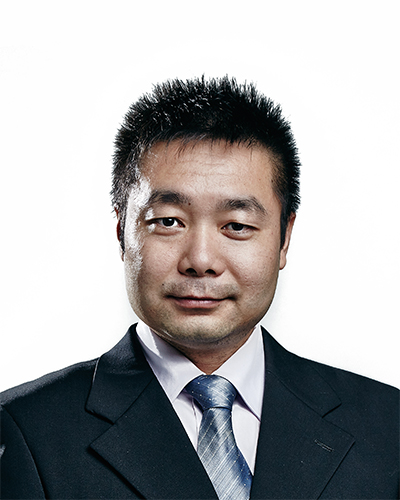}}]{Kai Yu}
is a professor at Computer Science and Engineering Department, Shanghai Jiao Tong University, China. He received his B.Eng. and M.Sc. from Tsinghua University, China in 1999 and 2002, respectively. He then joined the Machine Intelligence Lab at the Engineering Department at Cambridge University, U.K., where he obtained his Ph.D. degree in 2006. His main research interests lie in the area of speech-based human machine interaction including speech recognition, synthesis, language understanding and dialogue management. He is a member of the IEEE Speech and Language Processing Technical Committee.
\end{IEEEbiography}

\end{document}